\documentclass[12pt]{article}
\usepackage{epsfig}

\newcommand{\be}{\begin{eqnarray}}
\newcommand{\ee}{\end{eqnarray}}
\newcommand{\bea}{\begin{eqnarray}}
\newcommand{\eea}{\end{eqnarray}}

\newcommand{\beq}{\begin{equation}}
\newcommand{\eeq}{\end{equation}}

\newcommand{\nn}{\nonumber}

\def\la{\mathrel{\mathpalette\fun <}}

\def\fun#1#2{\lower3.6pt\vbox{\baselineskip0pt\lineskip.9pt
\ialign{$\mathsurround=0pt#1\hfil##\hfil$\crcr#2\crcr\sim\crcr}}}

\begin{document}

\title{Sigma-Meson and Confinement Singularity}

\author{A.V. Anisovich$^{1,2}$, V.V. Anisovich$^1$, M.A. Matveev$^1$,
K.V. Nikonov$^3$,\\ V.A. Nikonov$^{1,2}$,
J. Nyiri$^4$, A.V. Sarantsev$^{1,2}$ and T.O. Vulfs$^1$\\~\\
{\small $^1\,$Petersburg Nuclear Physics Institute, Gatchina,
Russia}\\
{\small $^2\,$Helmholtz-Institut f\"ur Strahlen- und Kernphysik,
Universit\"at Bonn, Germany}\\
{\small $^3\,$ Petersburg State University, Department of Physics,
St.Petersburg, Russia}\\
{\small $^4\,$Research Institute for Particle and Nuclear Physics,
Budapest, Hungary} }

\date{\today}

\maketitle

\begin{abstract}
We describe the meson-meson data for the ($IJ^{PC}=00^{++}$) wave at
$280\leq\sqrt s\leq 1900$ MeV in two approaches: (i) the K-matrix
approach and (ii) the dispersion relation D-matrix method.  With a
good description of low energy data (at $280\leq\sqrt s\leq 900$
MeV) as well as the data of two-meson transition amplitudes and
antiproton-proton annihilation into three pseudoscalar meson states
(at $450\leq\sqrt s\leq 1950$ MeV) we have found the positions of
the resonance poles: (i) for the $\sigma$ meson pole: $M_\sigma =
(390\pm 35)-i(235\pm 50)$ MeV; (ii) two poles for the $f_0(980)$, on
the second sheet (under the $\pi\pi$ cut): $M_{I} = (1011\pm
5)-i(35\pm 5)$ MeV, and on the third sheet (under the $\pi\pi$ and
$K\bar K$ cuts), $M_{II} = (1035\pm 50)-i(460\pm 50)$ MeV; for the
$f_0(1370)$ meson, $M= (1285\pm 30)-i(160\pm 20)$ MeV; for the
$f_0(1500)$ meson, $M = (1488\pm 4)-i(53\pm 5)$ MeV; for the
$f_0(1790)$ meson, $M = (1775\pm 25)-i(140\pm 15)$ MeV and for the
broad state $f_0(1200-1600)$ $M=(1540\pm 120)-i(550\pm 70)$ MeV. Our
estimation of the scalar-isoscalar scattering length obtained under
different parameterizations and assumptions about the quality of low
energy $\pi\pi$ scattering data is $a^0_0=(0.215\pm
0.040)\mu^{-1}_\pi$. We also discuss the idea according to which the
$\sigma$-meson could be a remnant of the confinement singularity,
$1/s^2$, in a white channel.
 \end{abstract}

\section{Introduction }

In spite of lengthy and persistent investigations, at present we
have no firm determination for the mass of the $\sigma$ meson - the
resonance in the 280--900 MeV region. This resonance reveals itself
in the $\pi\pi$ channel as a pole in the complex-$M$ plane, in the
$(IJ^{PC}=00^{++})$ partial wave. Numerous calculations produced
mass values distributed over all the low-energy interval
$\sqrt{s}\equiv M\la 900$ MeV, with various widths from 200 MeV up
to 1000 MeV. Such a situation emerged in the nineties \cite{PDG}.
The results of the latest analyses are clustered in a smaller mass
region 400-600 MeV: see, for example, \cite{Ablikim}
$(552^{+84}_{-106}) -i(232^{+81}_{-72})$ MeV and \cite{Garcia}
$(484\!\pm\!17) -i(255\!\pm\!10)$ MeV) and the review of Bugg
\cite{Bugg:2006gc} $472\!\pm\!30 - i(271\!\pm\!30)$ MeV. The
solution of the Roy equation at low energies produced a smaller mass
$441^{+16}_{-8} MeV-i(272^{+9}_{-13})$ MeV \cite{Caprini:2005zr}.

We see three sources for emerging uncertainties in the analyses
of the $\pi\pi$ amplitude near the threshold:\\
(i) a not sufficiently good determination of the $00^{++}$ amplitude
above $M=900$ MeV,\\
(ii) uncertainties in the definition of the left-hand cut in the
$\pi\pi$ amplitude and\\
(iii) uncertainties in low-energy $\pi\pi\to\pi\pi$ data.\\
In the present paper we analyzed in detail all these sources of
uncertainties. The examples considered in the paper demonstrate that
the results obtained for the low-energy amplitude depend strongly on
the assumptions made in the analysis.

The meson spectra in the $(00^{++})$ wave were fitted by our group
using the $K$-matrix technique \cite{km,kmR,APS}. This technique
provides us with an opportunity to fit simultaneously several
reactions (such as $\pi\pi,\,K\bar K,\,\eta\eta$, etc), taking into
account correctly analytical properties and unitarity in all
investigated amplitudes. This way we have determined the resonance
structure of the scalar-isoscalar wave at $500\leq \sqrt s\leq 1950$
MeV; our results were summarized in \cite{book3}.

However, in the $K$-matrix amplitude the left-hand cut owing to
crossing channels is determined ambiguously (note that $t$ and $u$
channel meson exchanges depend on couplings and form factors, which
are not well known). The impossibility to write down precisely the
contributions of left-hand cuts leads to a freedom in the
interpretation of the $\pi\pi\to \pi\pi$ amplitude in the
$\sqrt{s}<500$ MeV region. In our $K$-matrix analyses
\cite{km,kmR,APS} of the isoscalar scalar wave we modeled the
contribution from the left-hand cut at $s<0$ by introducing several
poles in this region with fitted parameters. Describing this partial
amplitude in the region $280\leq\sqrt{s}\leq 1900$ MeV, we usually
did not observe a pole which could be interpreted as the
$\sigma$-meson. However, in some solutions (not the best ones) such
a pole appeared.

Having this background, we fitted in \cite{AN} the amplitude
$00^{++}$ in the region $280\leq\sqrt{s}\leq 900$ MeV separately in
the framework of the dispersion relation approach sewing the
$N/D$-solution with the $K$-matrix one at $450\leq\sqrt{s}\leq 900$
MeV. As a result, the best fit, accounting for the left-hand cut
contribution (it was a fitting function), contained the
$\sigma$-meson pole at $M_{\sigma}=(430\pm 150) -i(320\pm 130)$ MeV
\cite{AN}.

One can think that the ambiguity problem may be solved with the help
of the investigation of the $\pi\pi$ scattering in all three
($u,d,s$) channels (see \cite{GarciaMartin:2011cn} and references
therein). However, this procedure requires the analytical
continuation of the pole terms into regions being rather far from
the pole mass. This supposes the knowledge not only of both
resonance form factors and the energy dependence of resonance
widths. The high spin states lead to the divergence in crossing
channels. It is only the summing over all sets of states that
resolves these divergences resulting finally in the Regge behavior
and therefore, requires model-dependent calculations.

\begin{figure}
\centerline{\epsfig{file=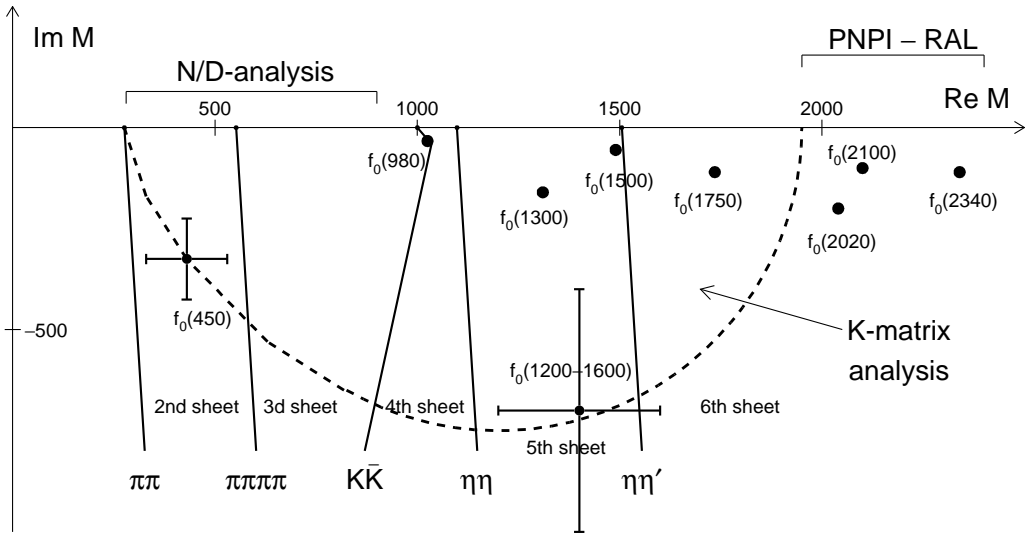,width=14cm}}
\caption{Complex-$M$ plane for the $(IJ^{PC}=00^{++})$ mesons
\protect\cite{book3}. The dashed line encircles the part of the
plane where the $K$-matrix analysis \protect\cite{kmR} reconstructs
the analytical $K$-matrix amplitude: in this area the poles
corresponding to resonances $f_0(980)$, $f_0(1300)$, $f_0(1500)$,
$f_0(1750)$ and the broad state $f_0(1200-1600)$ are located. Beyond
this area, in the low-mass region, the pole of the light
$\sigma$-meson is located (shown by the point the position of pole,
$M=(430-i320)$ MeV, corresponds to the result of $N/D$ analysis ;
the crossed bars stand for $\sigma$-meson pole found in
\protect\cite{AN}). In the high-mass region one has resonances
$f_0(2030),f_0(2100),f_0(2340)$, see \protect\cite{ufn}.  Solid
lines stand for the cuts related to the thresholds
$\pi\pi,\pi\pi\pi\pi,K\bar K,\eta\eta,\eta\eta'$.}
\label{Xpolesco2}
\end{figure}

The $K$-matrix analysis \cite{km,kmR,APS}, being performed at a
distance from the left-hand cut, gives masses and full widths of
resonances (i.e. the position of poles) as well as the residues of
the poles, namely, couplings of resonances to different channels.
These couplings are factorized; this is a criterium for dealing just
with a particle, though unstable. Besides, the coupling
interrelations allow one to define the quark content of a particle,
provided this is a $q\bar q$ state. This way the states found in the
$K$-matrix analysis can be classified as $q\bar q$ nonets.
The $K$-matrix analysis determines two nonets and one extra state
 in the 600-2000 MeV region. One of the possible classifications is
 given in \cite{book3}:
$$[f_0(980),f_0(1300)]_{n=1},\quad [f_0(1500),f_0(1750)]_{n=2}\, ;$$
where $n=1,2$ are the radial quantum numbers. Here the broad state
 $f_0(1200-1600)$ and the $\sigma$-meson are superfluous
 for the $q\bar q$ nonet classification.
The position of resonances in the $IJ^{PC}$=$00^{++}$ wave is shown
in Fig. \ref{Xpolesco2}.

In this classification the broad state $f_0(1200-1600)$ is a
glueball descendant \cite{glueball,ZPhys}. Due to another
classification the broad $f_0(1200-1600)$ state belongs to the first
nonet and the extra state is $f_0(1300)$. Both these states are
flavor blind and one of them is superfluous for the $q\bar q$
systematics. The $f_0(1200-1600)$ state acquired a large width
because of the accumulation of widths of neighboring states: in
nuclear physics such a phenomenon had been studied in
\cite{Shapiro,Okun,Stodolsky}, in meson physics in \cite{ABS}.



The paper is organized as follows. In Section 2 we provide formulae
used in the $K$-matrix and $D$-matrix approaches. In Section 3 we
discuss an idea of the confinement singularity $1/s^2$. Such a
singularity in the t-channel ($1/t^2$) corresponds to the linear
rising potential which describes meson spectra in $q\bar q$
\cite{SI-qq}, $b\bar b$ \cite{SI-bb} and $c\bar c$ \cite{SI-cc}
channels and gives correct values for the partial widths of
radiative and hadronic decays of confined $q\bar q$ states
\cite{gpi}. Although this singularity is expected to be in the color
octet state, it can have also a color singlet component and appear
in the s-channel. The $K$-matrix and $D$-matrix analyses of the
$00^{++}$ wave in the energy interval $280\leq\sqrt{s}\leq 1900$ MeV
are presented in Section 4. In the Conclusion we summarize the
results concentrating on the low-energy region.

Some clarifying points are made in the Appendices. In Appendix A the
dynamical mechanism of the singularity $1/s^2$ in the $q\bar q\to
q\bar q$ amplitude is discussed. A simple description of the low
energy $\pi\pi$ scattering ($270\leq \sqrt s \leq 900$ MeV) in terms
of the dispersion relation approach which allows to incorporate
easily the singularity $1/s^2$ into the analytic and unitary
amplitudes is given in Appendix B. In Appendix C we present the
unitary $\pi\pi$ scattering in the threshold regions taking into
account the mass differences of the $\pi^+\pi^-$ and $\pi^0\pi^0$
systems which are essential for the extraction of $a^0_0$.

\section{The $K$-Matrix and $D$-Matrix Techniques}

Here we discuss the analytic properties of amplitudes restored in
terms of the $K$-matrix and $D$-matrix techniques.

\subsection{The $K$-matrix approach}

For the $S$-wave interaction in the isoscalar sector we use, as
previously \cite{kmR}, the 5-channel $K$-matrix:
\be
 K_{ab}^{00}(s)
=\left ( \sum_\alpha \frac{g^{(\alpha)}_a g^{(\alpha)}_b}
{M^2_\alpha-s}+f_{ab} \frac{1\;\mbox{GeV}^2+s_0}{s+s_0} \right )\;
\frac{s-s_A}{s+s_{A0}}\;\;,
 \label{km_elem}
\ee
 where $K_{ab}^{IJ}$ is a 5$\times $5 matrix ($a,b$ = 1,2,3,4,5),
with the following notations for meson states: 1 = $\pi\pi$, 2 =
$K\bar K$, 3 = $\eta\eta$, 4 = $\eta\eta'$ and 5 =  multimeson
states (four-pion state mainly at $\sqrt{s}<1.6\; \mbox{GeV}$). The
$g^{(\alpha)}_a$ are coupling constants of the bare state $\alpha$
to meson channels; the parameters $f_{ab}$ and $s_0$ describe the
smooth part of the $K$-matrix elements ($s_0>1.5$ GeV$^2$). The
factor $(s-s_A)/(s+s_{A0})$, where $s_A\sim (0.1-0.5)m^2_\pi$,
 describes Adler's zero in the two-pion channel. However, in the
K-matrix analysis we introduced this factor also in other channels
to suppress the effect of the left-hand side false kinematic
singularities in the $K$-matrix amplitude.

\begin{figure}[h]
\centerline{\epsfig{file=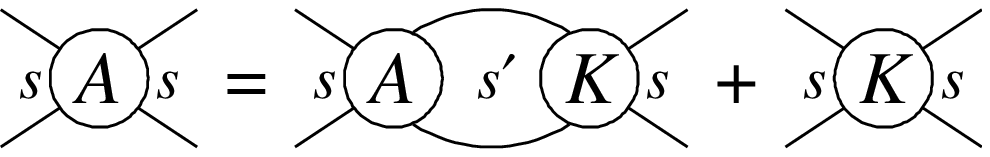,width=9cm}}
\caption{Graphical representation of the spectral
integral equation for the K-matrix amplitude.
\label{BS_amp}}
\end{figure}

\subsubsection{Spectral
integral equation for the K-matrix amplitude}

Discussing meson-meson scattering and production amplitudes, we use
the dispersion relation (or spectral integral) technique. In terms
of this technique we write for the K-matrix amplitude a spectral
integral equation which is an analog of the Bethe-Salpeter equation
\cite{Salpeter} for the Feynman technique. The spectral integral
equation for the transition amplitude from the channel $a$ to
channel $b$ is presented graphically in Fig.~\ref{BS_amp} and reads:
 \be A_{ab}(s)=
\int\frac{ds'}{\pi}\frac{A_{aj}(s,s')}{s'-s-i0}\rho_j(s')
K_{jb}(s',s)+K_{ab}(s)\ .
\label{bethe}
\ee
 Here $\rho_j(s')$ is the diagonal matrix of the phase volumes,
$A_{aj}(s,s')$ is the off-shell amplitude and $K_{jb}(s,s')$ is the
off-shell elementary interaction. Let us remind that in the
dispersion relation technique, just as in quantum mechanics, there
is no energy conservation for the intermediate states. For more
details see \cite{book3}, Chapter 3.

The standard way of the transformation of eq. (\ref{bethe}) into the
K-matrix form  is the extraction of the imaginary and principal
parts of the integral. The principal part has no singularities in
the physical region and can be omitted (or taken into account by a
re-normalization of the K-matrix parameters):
\bea
\int\frac{ds'}{\pi}\frac{A_{aj}(s,s')}{s'-s-i0}\rho_j(s')
K_{jb}(s',s)&=&
P\int\frac{ds'}{\pi}\frac{A_{aj}(s,s')}{s'-s}\rho_j(s')
K_{jb}(s',s)+iA_{aj}(s,s)\rho_j(s)K_{jb}(s)\nn \\
&\to& iA_{aj}(s,s)\rho_j(s)K_{jb}(s)
\label{bethe2}
\eea
 For the amplitude $A_{ab}(s)$ one obtains the standard
$K$-matrix expression which in the matrix form reads:
\be
 \hat A=\hat A\,i\,\hat \rho \hat K+\hat K, \quad {\rm or}\qquad \hat
A=\hat K(I-i\hat \rho\hat K)^{-1}\ ,
\label{km}
\ee
 The factor $(I-i\hat \rho\hat K)^{-1}$ describes the rescattering
of mesons, it is inherent not only in two-meson transition
amplitudes but in production amplitudes as well. The $P$-vector
method describes the production of particles in cases when an
initial interaction should be taken into account only
 once, for example, for the production of mesons from the
$\gamma\gamma$ collision or from proton-antiproton annihilation:
\be
A_k(\bar p p)=P_j\left [(I-i\hat \rho\hat K)^{-1}\right ]_{jk}\, .
\label{kmP1}
\ee
 Elements of the vector $P_j$ have a form similar to
 the K-matrix elements, eq. (\ref{km_elem}):
\be
P_j= \sum_\alpha \frac{\Lambda_\alpha g^{(\alpha)}_j}
{M^2_\alpha-s}+F_{j}\;.
\label{kmP2}
\ee
The first term in eq.(\ref{kmP2}) refers to the production of
resonances; the second one, $F_{j}$, to a non-resonant
production.

The standard form of the two-particle phase volume is
\beq
\rho_a(s,m_{1a},m_{2a})=\sqrt{\frac{(s-(m_{1a}+m_{2a})^2)(s-(m_{1a}-m_{2a})^2)}{s^2}},
\quad a=1,2,3,4
\label{3}
\eeq
 where $m_{1a}$ and $m_{2a}$ are masses of the final particles.
In the case of different masses this expression includes the term
$\sqrt{s-(m_{1a}-m_{2a})^2}$ which in the K-matrix approach can be a
source of false kinematic singularities on the first (physical)
sheet: the loop diagram amplitude, $B(s)$, does not contain this
type of singularities. Such a cancelation can be taken into account
effectively by replacing the $\eta\eta'$ phase volume:
\be
\sqrt{\frac{(s-(m_{1a}+m_{2a})^2)(s-(m_{1a}-m_{2a})^2)}{s^2}}
\to\sqrt{\frac{s-(m_{1a}+m_{2a})^2}{s}}\, .
\label{4}
\ee

For the restoration of the amplitude we need to take into account
not only the cuts related to threshold singularities of the stable
particles but non-stable ones as well. In the $00^{++}$-amplitude
the four-pion state gives cuts related to $\pi\pi\pi\pi$ (at the
real $s$-axis, $\sqrt{s} =4m_\pi$) and in the complex-$s$ plane
related to the production of vector and scalar particles:
$\pi\pi\rho$ (at $\sqrt{s} =2m_\pi+m_\rho$ with a complex mass
$m_\rho$), $\rho\rho$ (at $\sqrt{s} =2m_\rho$) and $f_0f_0$. Let us
write the phase space factor for the $\rho\rho$-state which contains
$4\pi$, $\pi\pi\rho$ and $\rho\rho$ threshold singularities:
\bea
&&\rho_{4\pi}(s)=\int\limits_{4\,m_\pi^2}^{(\sqrt{s}-2m_\pi)^2}
\frac{ds_{12}}{\pi}
\int\limits_{4\,m_\pi^2}^{(\sqrt{s}-\sqrt{s_{12}})^2}
\frac{ds_{34}}{\pi} G^2_{in}(s,s_{12},s_{34})\,
\rho(s,\sqrt{s_{12}},\sqrt{s_{34}})\times\nn\\
&&\frac{G^2(s_{12})(s_{12}\!-\!4\,m^2_\pi)\rho(s_{12},m_\pi,m_\pi)}
{(s_{12}-M^2_\rho)^2+(M_\rho\Gamma_\rho)^2}\,
\frac{G^2(s_{34})\,(s_{34}\,-\,4\,m^2_\pi)\rho(s_{34},m_\pi,m_\pi)}
{(s_{34}-M^2_\rho)^2+(M_\rho\Gamma_\rho)^2}
\label{4pi}
\eea
The form factors $G_{in}(s,s_{12},s_{34})$, $G(s_{12})(s_{12})$,
 $G(s_{34})$ are introduced into (\ref{4pi}) to provide the
convergency of the integrals. This phase volume describes production
of $\rho\rho$ in the $S$-wave and $P$-wave production of pions in
the $\rho$-meson decays. Being near a pole, hadronic production cuts
split this pole into several ones located on different sheets of the
complex-$s$ plane.

\subsection{The $D$-matrix approach}

The considered above approaches allow us to distinguish between
``bare'' and ``dressed'' particles: due to meson rescattering the
bare particles, with poles on the real-$s$ axis, are transformed
into particles dressed by ``coats'' of meson states. In the K-matrix
approach we deal with a ``coat'' formed by real particles -- the
contribution of virtual ones is included in the principal part of
the loop diagram, $B(s)$, and is taken into account effectively by
the re-normalization of mass and couplings.

In the dispersion relation D-matrix approach one can take into
account the ``coat'' of virtual mesons. The D-matrix amplitudes
describe transitions of bare states.

Let us consider the block $D_{\alpha\beta}$ which describes a
transition between the bare state $\alpha$ (but without the
propagator of this state) and the bare state $\beta$ (with the
propagator of this state included). For such a block one can write
the following equation:
\be
D_{\alpha\beta}= D_{\alpha\gamma}\sum\limits_j
B^j_{\gamma\eta}d_{\eta\beta}+d_{\alpha\beta}
\ee
 Or, in the matrix form:
 \be
 \hat D= \hat D\hat B\hat d+\hat d \qquad \hat D= \hat
d(I-\hat B\hat d)^{-1}
 \ee
  Here the $\hat d$ is a diagonal matrix of the propagators:
 \be
 \hat d=diag\left
(\frac{1}{M^2_1-s},\frac{1}{M^2_2-s},\ldots,\frac{1}{M^2_N-s},R_1,R_2
\ldots\right)
\ee
 where $R_\alpha$ are propagators for non-resonant
transitions (discussed below), and the elements of the $\hat
B$-matrix are equal to:
 \be \hat B_{\alpha\beta}=\sum\limits_j
B^j_{\alpha\beta}=
\sum\limits_j\int\frac{ds'}{\pi}\frac{g^{R(\alpha)}_j\rho_j(s',m_{1j},m_{2j})g^{L(\beta)}_j}{s'-s-i0}
\;.
 \ee
  The $g^{R(\alpha)}_j$ and $g^{L(\alpha)}_j$ are right and
left vertices for a transition from the bare state $\alpha$ to the
channel $j$. For the pole terms there is a clear factorization:
\be
g^{R(\alpha)}_j=g^{L(\alpha)}_j=g^{(\alpha)}_j\,. \label{ver_pole}
\ee

However, non-resonant terms do not provide such a factorization. A
solution of this problem is to introduce for non-resonant
transitions a separate propagator and vertices from every initial
state $i$. Moreover, for the description of the non-resonant terms
between different initial and final states a second propagator with
permutated left and right vertices is needed. In this case the
propagator index provides automatically a unique identification of
the transition term. Then for non-resonant transitions from the
$\pi\pi$ channel we have:
\be g_i^{L(N+1)}R_1
g_j^{R(N+1)}+g_i^{L(N+2)}R_2 g_j^{R(N+2)}
\ee
where $N$ is the
number of pole terms. The non-zero left and right vertices can be
taken as:
\bea
&&g_j^{L(N+1)}=f_{1j}\frac{1\;\mbox{GeV}^2+s_0}{s+s_0} \qquad
g_1^{R(N+1)}=1\qquad R_1=1\,, \nonumber \\
&&g_{1}^{L(N+2)}=1 \qquad
g_{j>1}^{R(N+2)}=f_{1j}\frac{1\;\mbox{GeV}^2+s_0}{s+s_0}  \qquad
R_2=1
\label{case_1}
\eea
and
\be
g_{j>1}^{R(N+1)}=g_{j>1}^{L(N+2)}=g_{1}^{R(N+2)}=0
\ee

Another alternative parametrization for the non-zero terms is:
\be
&&g_j^{L(N+1)}=f_{1j}\qquad g_1^{R(N+1)}=1\qquad R_1=
\frac{1\;\mbox{GeV}^2+s_0}{s+s_0}\,, \nonumber \\
&&g_{1}^{L(N+2)}=1 \qquad g_{j>1}^{R(N+2)}=f_{1j}\qquad
R_2=\frac{1\;\mbox{GeV}^2+s_0}{s+s_0}\,.
\label{case_2}
\ee

With such a definition the amplitude $A_{ab}$ is the convolution of
the matrix $D_{\alpha\beta}$ with right and left coupling vectors,
$g^{(R,\alpha )}_a$ and $g^{(L,\beta)}_b$:
\be
A_{ab}=\sum\limits_{\alpha,\beta} g^{R(\alpha )}_a d_{\alpha\alpha}
D_{\alpha\beta}g^{L(\beta)}_b\,.
\ee
The P-vector amplitude has the form:
\be
A_{b}=\sum\limits_{\alpha,\beta} \tilde P^{(\alpha )}
d_{\alpha\alpha} D_{\alpha\beta}g^{L(\beta)}_b\qquad \tilde P=\left
(\Lambda_1,\Lambda_2,\ldots,\Lambda_n,F_1/R_1\ldots\right )
\ee
where couplings $\Lambda_\alpha$ and non-resonant terms $F_j$ are
the same as in eq.(\ref{kmP2}).

In the present fits we calculate the elements of the
$B^j_{\alpha\beta}$ using one subtraction taken at the channel
threshold $M_{j}=(m_{1j}+m_{2j})$:
\be
B^{j}_{\alpha\beta}(s)=B^j_{\alpha\beta}(M_{j}^2)+(s-M_j^2)
\int\limits_{m_a^2}^\infty \frac{ds'}{\pi}
\frac{g^{R(\alpha)}_j\rho_j(s',m_{1j},m_{2j})g^{L(\beta)}_j}{(s'-s-i0)(s'-M^2_{j})}.
\label{D9} \ee
In the case of the non-resonant terms parameterized in the form
(\ref{case_2}) and the S-wave vertices parameterized as constants
the expression for elements of the $\hat B$ matrix can be rewritten
as:
\be
B^{j}_{\alpha\beta}(s)=g^{R(\alpha)}_a\left ( b^j+(s-M_j^2)
\int\limits_{m_a^2}^\infty \frac{ds'}{\pi}
\frac{\rho_j(s',m_{1a},m_{2a})}{(s'-s-i0)(s'-M^2_{j})}\right )
g^{L(\beta)}_b\,,
\label{D9a} \ee
where the parameters $b^j$ depend on decay channels only.

\begin{table}[h]
\begin{center}
\caption{\label{decay}
Coupling constants given by quark
combinatorics for $(q\bar q)_{I=0}$ meson and glueball decays into two
pseudoscalar mesons in the leading terms of the $1/N_c$ expansion. The
$\Phi$ is the mixing angle for $n\bar n=(u\bar u+d\bar d)/\sqrt 2$ and
$s\bar s$ states: $n\bar n\cos\Phi+s\bar s\sin\Phi$. The $\Theta$ is
the mixing angle for $\eta -\eta'$ mesons:
$\eta=n\bar n \cos\Theta-s\bar s \sin\Theta$ and
$\eta'=n\bar n \sin\Theta+s\bar s \cos\Theta$ with
$\Theta\simeq 37^o$ }
\vskip 0.5cm
\begin{tabular}{c|c|c|c} decay      &$q\bar q$-meson decay coupling&
$gg$ state decay coupling &identity \\
channel&~&~&factor   \\
~&~&~&~\\
 $\pi^0\pi^0$ & $g\;\cos\Phi/\sqrt{2}$&$G$&
1/2  \\ $\pi^+\pi^-$ & $g\;\cos\Phi/\sqrt{2}$ &$G$&  1   \\
$K^+K^-$ & $g
(\sqrt 2\sin\Phi+\sqrt \lambda\cos\Phi)/\sqrt 8 $ &$\sqrt{\lambda} G$&
1 \\
$K^0\bar K^0$ & $g (\sqrt 2\sin\Phi+\sqrt{\lambda}\cos\Phi)/\sqrt 8 $
&$\sqrt \lambda G$& 1 \\
$\eta\eta$ & $g(\cos^2\Theta\;\cos\Phi/\sqrt 2
+\sqrt{\lambda}\;\sin\Phi\;\sin^2\Theta$)
 &$G(\cos^2\Theta+\sqrt{\lambda} \sin^2\Theta)$& 1/2 \\
$\eta\eta'$ &
$g\sin\Theta\;\cos\Theta(\cos\Phi/\sqrt 2-\sqrt{\lambda}\;\sin\Phi$) &
$G(1-\lambda)\cos\Theta\sin\Theta$&1
\end{tabular}
\end{center}
\end{table}

In the case of the D-matrix approach it is not needed to introduce
the regularization of the $\eta\eta'$ phase volume and, therefore,
we use the standard expression (\ref{3}). It is also not necessary
to introduce any regularization for the D-matrix elements at $s=0$:
this point is not singular in this approach. Thus, in the D-matrix
fits, the term with the Adler zero was introduced in the $\pi\pi$
channel only. Technically, it can be done either by the modification
of vertices or by the modification of the $\pi\pi$ phase volume: \be
\rho_1(s,m_\pi,m_\pi)=\frac{s-s_A}{s+s_{A0}}\sqrt{\frac{s-4m_\pi}{s}}
\ee

For $q\bar q$ states one can relate the decay couplings
$g^{(\alpha)}_a$ in terms of the rules of quark combinatorics (see
\cite{book3}, Chapter 2, and references therein). The couplings for
channels $a=\pi\pi$, $K\bar K$, $\eta\eta$, $\eta\eta'$, calculated in
leading terms of the $1/N_c$ expansion, are presented in Table
\ref{decay}. The couplings depend on the constant $g$ which is
universal for all nonet states, the mixing angle $\Phi$ which
determines the proportion of the $n\bar n=(u\bar u+d\bar d)/\sqrt 2$
and $s\bar s$ components in the decaying $q\bar q$ state, and the
$s\bar s$ production suppression parameter $\lambda\sim 0.5-0.7$.
Two scalar-isoscalar states of the same nonet are orthogonal if:
\be
\Phi^{(I)}-\Phi^{(II)}=\pm 90^o.
\label{D12}
\ee
 The equality of the coupling constants $g$ and the fulfilment of the
mixing angle relation (\ref{D12}) is a basis for the determination
of mesons of a $q\bar q$-nonet.

 The gluonic states are
decaying in the channels $a=\pi\pi,\,K\bar K,\,\eta\eta,\eta\eta'$
with the same couplings as the $q\bar q$-state but at a fixed mixing
angle $\Phi\to \Phi_{glueball}$ which is determined by the value of
$\lambda$, namely: $\Phi_{glueball}=\cos^{-1}\sqrt{2/(2+\lambda)}$.
The corresponding couplings are given in Table \ref{decay} as well.

\section{Confinement Interaction in the $q\bar q$ Sector}

The description of  mesons of the $q\bar q$ sector is a source of
information about quark confinement interaction. These interactions
contain $t$-channel singularities of scalar and vector type. The
$t$-channel exchange interaction can be both in white and colour
states, ${\bf c}={\bf 1}+{\bf 8}$ though, of course, the colour-octet
interaction plays a dominant role in meson formation.

The observed linearity of the $q\bar q$-meson trajectories in the
($n,M^2$) planes \cite{syst}, where $n$ is the radial quantum number
of the $q\bar q$-meson with mass $M$, provides us the $t$-channel
singularity $V_{conf}\sim 1/q^4$ or, in coordinate representation,
$V_{conf}\sim r$. In the coordinate representation the confinement
interaction can be written in the following potential form
\cite{book3,SI-qq}:
 \bea \label{I1}
&&V_{conf}=(I\otimes I)\,b_S\,r +
  (\gamma_\mu\otimes \gamma_\mu)\,b_V\,r\ , \\ \nn &&b_S\simeq -b_V
  \simeq 0.15\,\, {\rm GeV}^{-2}\ .
\eea
 The first term in (\ref{I1}) refers to scalar interaction ($I\otimes I$),
the second one to vector ($\gamma_\mu\otimes \gamma_\mu$) - in the
$q\bar q$ sector the scalar and vector  forces are approximately equal.

\subsection{White remnants of the confinement singularities}

\begin{figure}
\centerline{\hspace{6mm}\epsfig{file=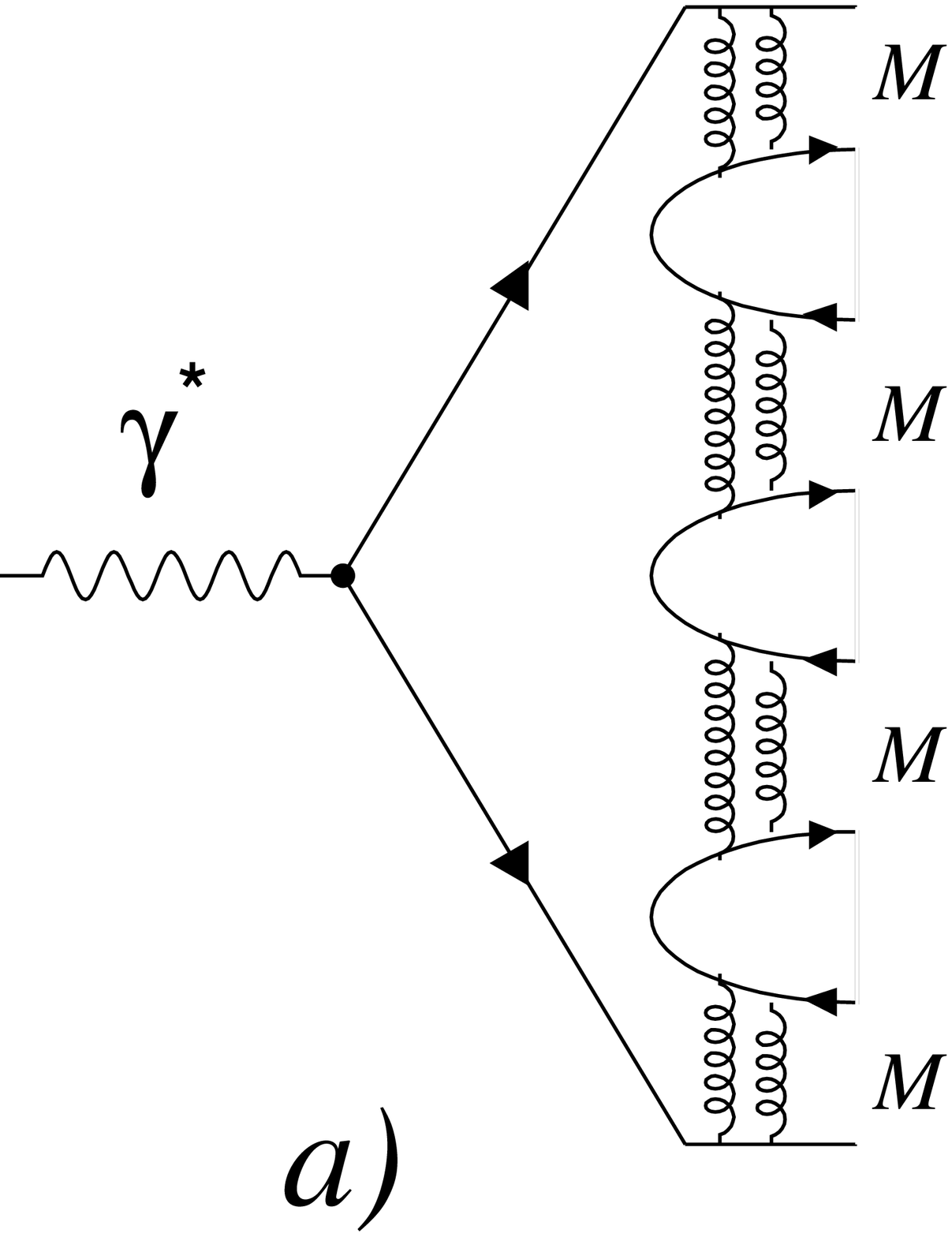,height=4cm}
            \epsfig{file=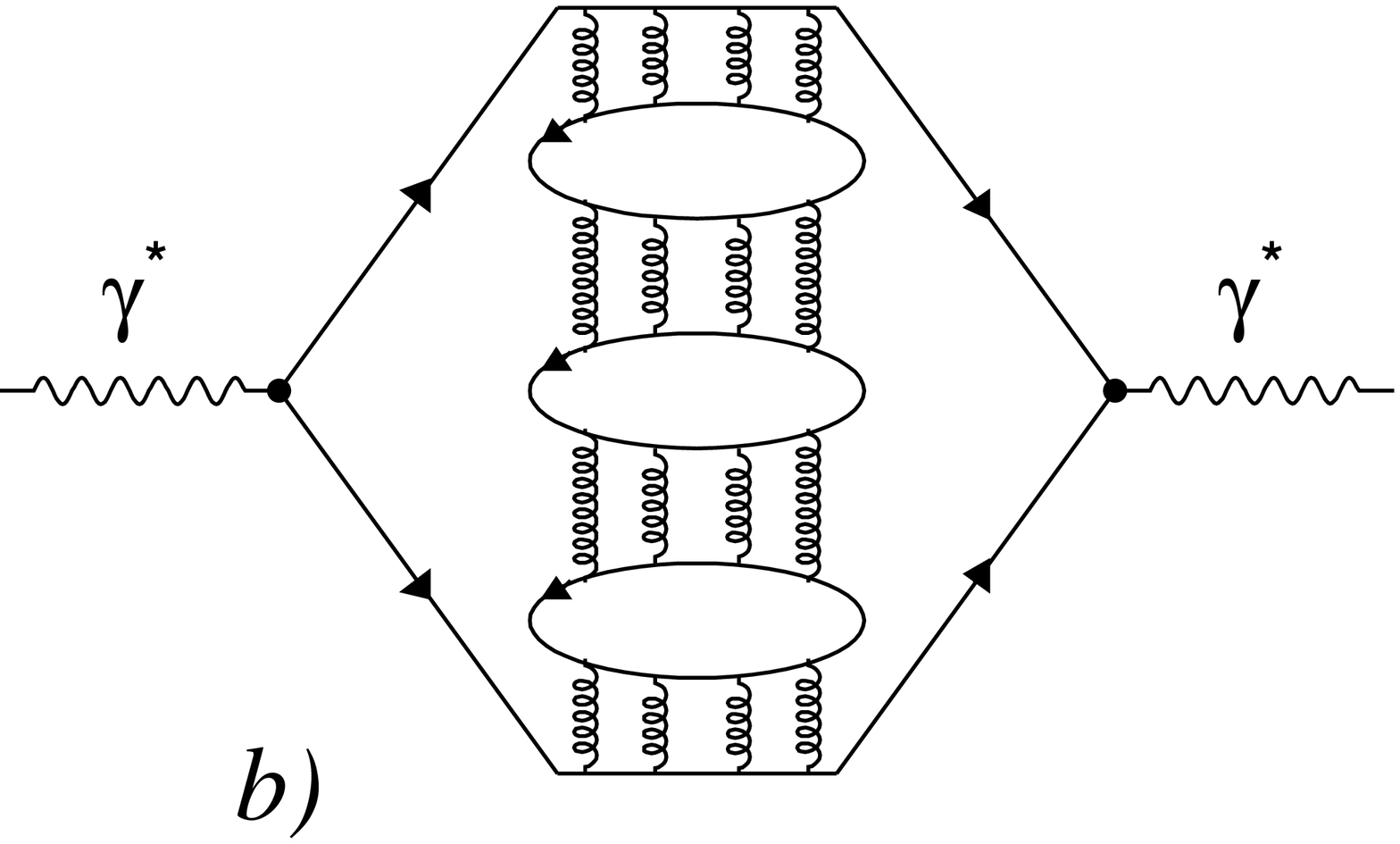,height=4cm}
            \epsfig{file=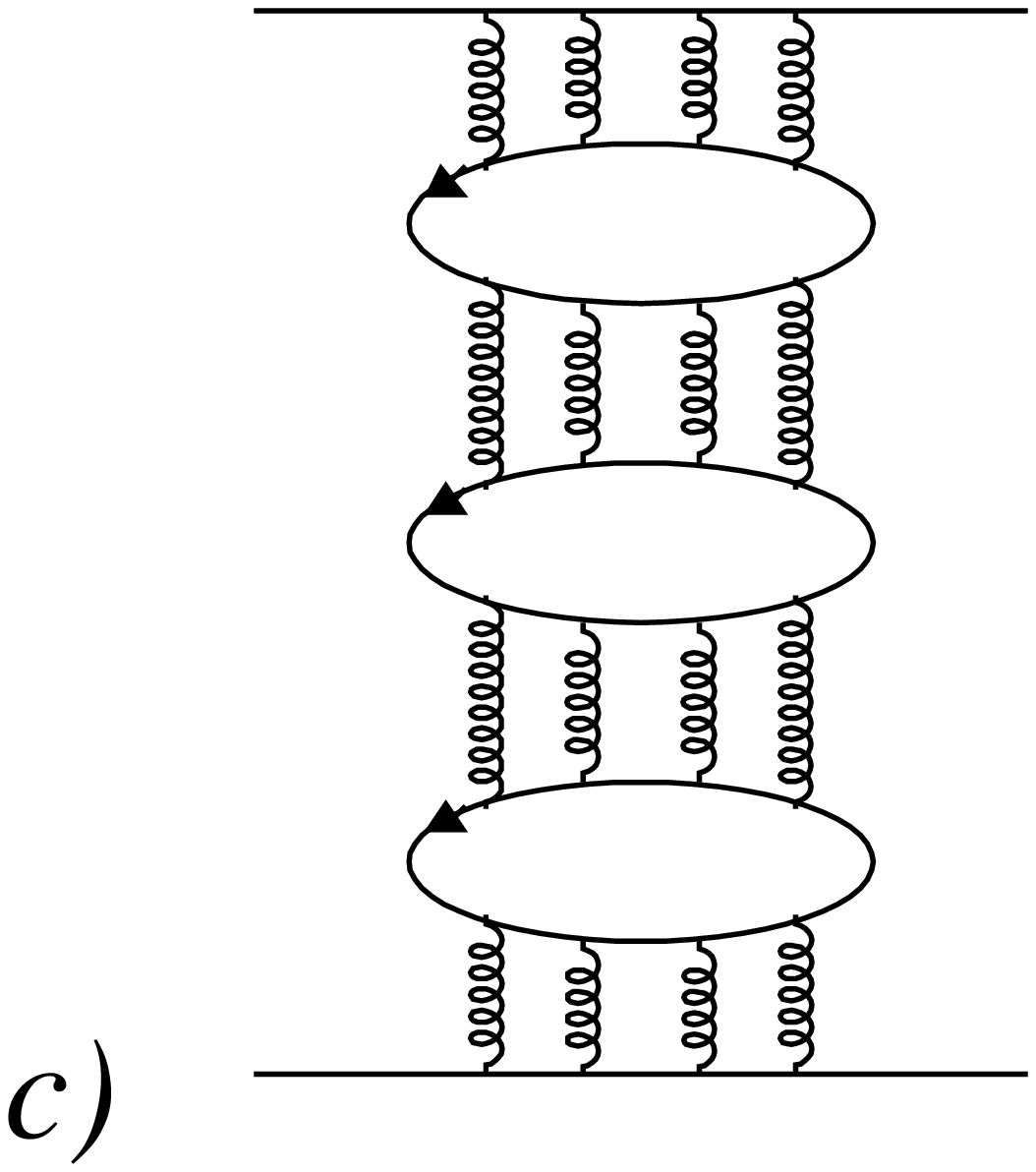,height=4cm}}
\caption{a) Quark--gluonic comb produced by breaking  a string by
quarks flowing out in  the process $e^+e^- \to \gamma^*\to q\bar
q\to mesons$.  b) Convolution of the quark--gluonic combs. c)
Example of diagrams describing interaction forces in the $q\bar q$
systems.}
\label{X2f22}
\end{figure}

We have serious reasons to suspect that the confinement
singularities (the $t$-channel singularities in the scalar and
vector states) have a complicated structure. In the color space
these are octet states but, may be, they contain also white
components. The octet exchange interaction contains quark-antiquark
and gluonic blocks. Therefore, the question is whether
$V^{(1)}_{confinement}(q^2)$ has the same singular behavior as
$V^{(8)}_{confinement}(q^2)$. The observed linearity of the
$(n,M^2)$-trajectories, up to the large-mass region,
$M\sim2000-2500$ MeV \cite{syst},  favors the idea of the
universality in the behavior of potentials $V^{(1)}_{confinement}$
and $V^{(8)}_{confinement}$ at large $r$, or small $q$. To see that,
let us consider, as an example, the process $\gamma^*\to q\bar q$,
Fig. \ref{X2f22}a.  We discuss the color neutralization mechanism of
outgoing quarks as a breaking of the gluonic string by newly born
$q\bar q$-pairs, see the discussion in \cite{Gribov}. At
 large distances, which correspond to
the formation of states with large masses, several new $q\bar
q$-pairs should be formed. It is natural to suggest that a
convolution of the quark--gluon combs governs the interaction forces
of quarks at large distances, see Fig. \ref{X2f22}b. The mechanism
of the formation of new $q\bar q$-pairs to neutralize color charges
does not have a selected color component. In this case all color
components $3\otimes\bar3=1+8$ behave similarly, that is, at small
$q^2$ the singlet and octet components of the potential are
uniformly singular, $V^{(1)}_{confinement}(q^2)\sim
V^{(8)}_{confinement}(q^2) \sim1/q^4$.

If the confinement singularities have, indeed, white constituents,
this raises immediately the following questions:\\
(i) How do these constituents reveal themselves in white channels?\\
(ii) Can they be identified?\\
In the scalar channel we face the problem of the $\sigma$ meson
$(IJ^{PC}=00^{++})$: what is the nature of this state? If the white
scalar confinement singularity exists, it would be reasonable to
consider it as the $\sigma$ meson revealing itself: because of the
transitions into the $\pi\pi$ state, the confinement singularity
could move to the second sheet. If so, the $\sigma$ meson can
certainly not reveal itself as a lonely amplitude singularity
$1/t^2$ but a standard amplitude pole or a group of poles.

A similar scenario may be valid also for the vector confinement
singularity in the $\pi\pi\pi$ $(IJ^{PC}=01^{--})$ channel. In this
case it is natural to assume that the white confinement singularity
couples with the channel $\rho\pi$, splits  and  dives into the
complex-$M_{\pi\pi\pi}$ plane.

An illustrative example of a set of loop diagrams of the Fig.
\ref{X2f22}c type is considered in Appendix A. In this example we
demonstrate how the strong singularity, $1/t^2$, may arise in scalar
and vector channels of the interaction block. An analysis of the
simple case when we have only two poles on the lower part of the
second $s$-plane is performed in Appendix B: the obtained result is
in agreement with those obtained in analyses performed over
low-energy data only \cite{W800P,W800E,W800A,W800Ishida}.

\section{The $K$-Matrix and $D$-Matrix Approaches in Fits to the Data
at $0.28\leq\sqrt{s}\leq 1.95$ GeV}

Here we present a comparative analysis of the results obtained with
the $K$-matrix and $D$-matrix methods. These approaches give rather
similar results for the $f_0$-resonances at $\sqrt{s}\leq 2$ GeV. In
Table~\ref{table_chi} we show the data used in these analyses and
give corresponding $\chi^2$ for different fits. In
Table~\ref{table_const} we list the masses of bare states, mixing
angles and other parameters used in the minimization procedure.

\subsection{The $K$-matrix fit}

In the analysis of the present data set we fitted data in two steps.
In the first step all couplings were optimized as free parameters;
in the second step we imposed relations Table~\ref{decay} for the
poles with masses above 1 GeV. We did not observe any deterioration
of the data description due to these restrictions but a rather
notable improvement in the convergency of the fits. For the lowest
K-matrix pole we do not impose any constraints: the global coupling
and mixing angle for this pole given in Table~\ref{table_const} are
simply calculated from the couplings into the $\pi\pi $ and
$K\bar K$ channels.

In the present solutions there are two candidates for a glueball: it
is either the third or the fourth K-matrix pole (with a mass around
1200 MeV). For the glueball candidate we introduced in addition a
glueball decay coupling (see Table~\ref{decay}). However, this coupling
provided only a small improvement and did not allow us to
distinguish between these two cases.

The fit is hardly sensitive to the $\pi\pi\pi\pi$ couplings for the
two lowest K-matrix poles; in the final solution we fix them to be
zero.

To get a combined description of all reactions, we introduced
non-resonant terms for the transition from the $\pi\pi$ channel to
other final states. We did not find a notable sensitivity to
non-resonant transitions between other channels.

The $K_{e4}$ data can be described with a very small re-optimization
of the K-matrix parameters found in \cite{km,kmR,APS}. We did not
find any change in the pole structure of the scalar-isoscalar
amplitude above 900 MeV. However, one of the pole singularities
situated around $s=0$ moved to higher masses. Its position, as well
as the positions of other poles, is given below in
(\ref{table_pole}).

\subsection{The $D$-matrix fits}

\begin{figure}[h]
\begin{center}
\epsfig{file=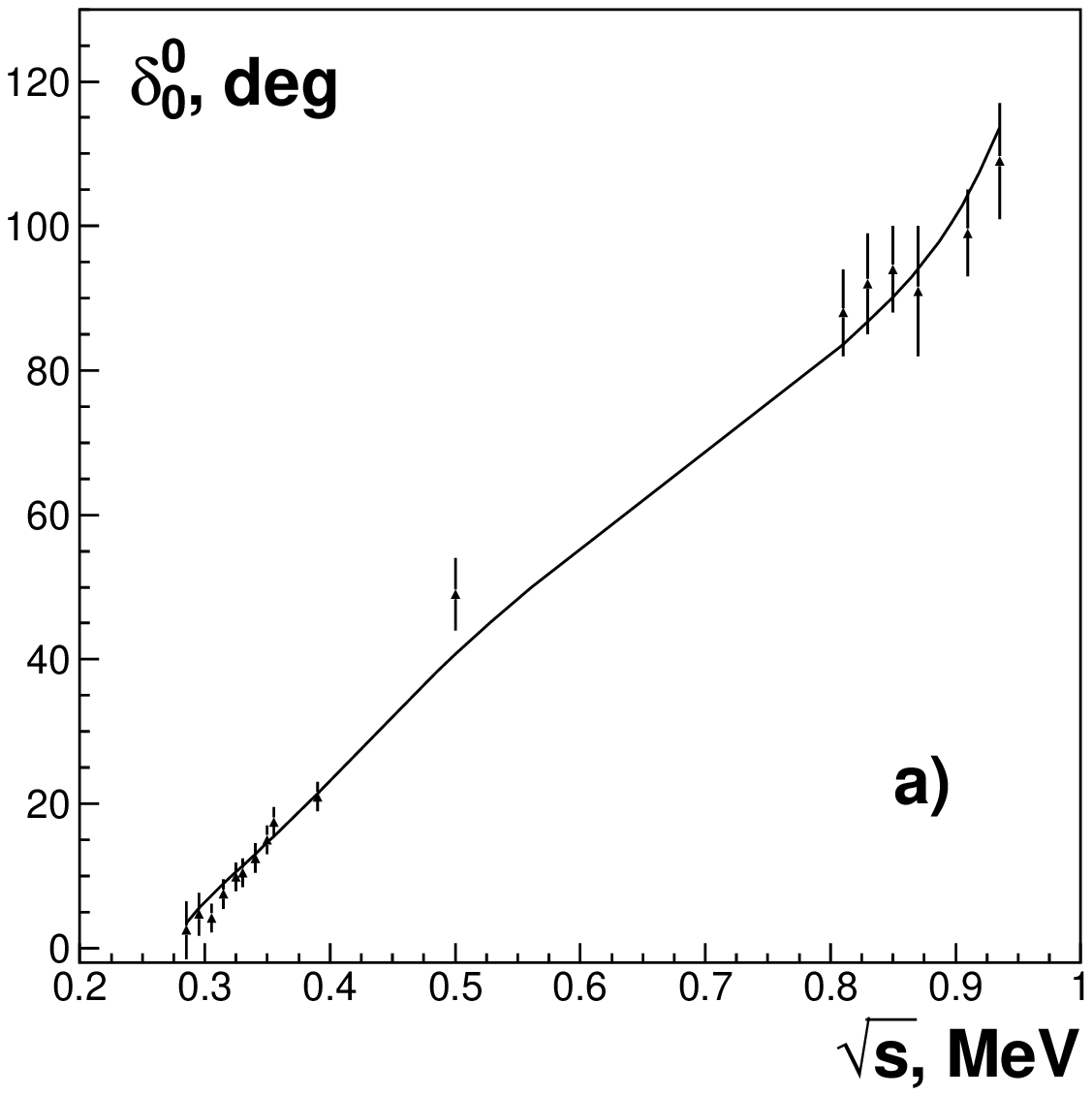,width=6cm}
\epsfig{file=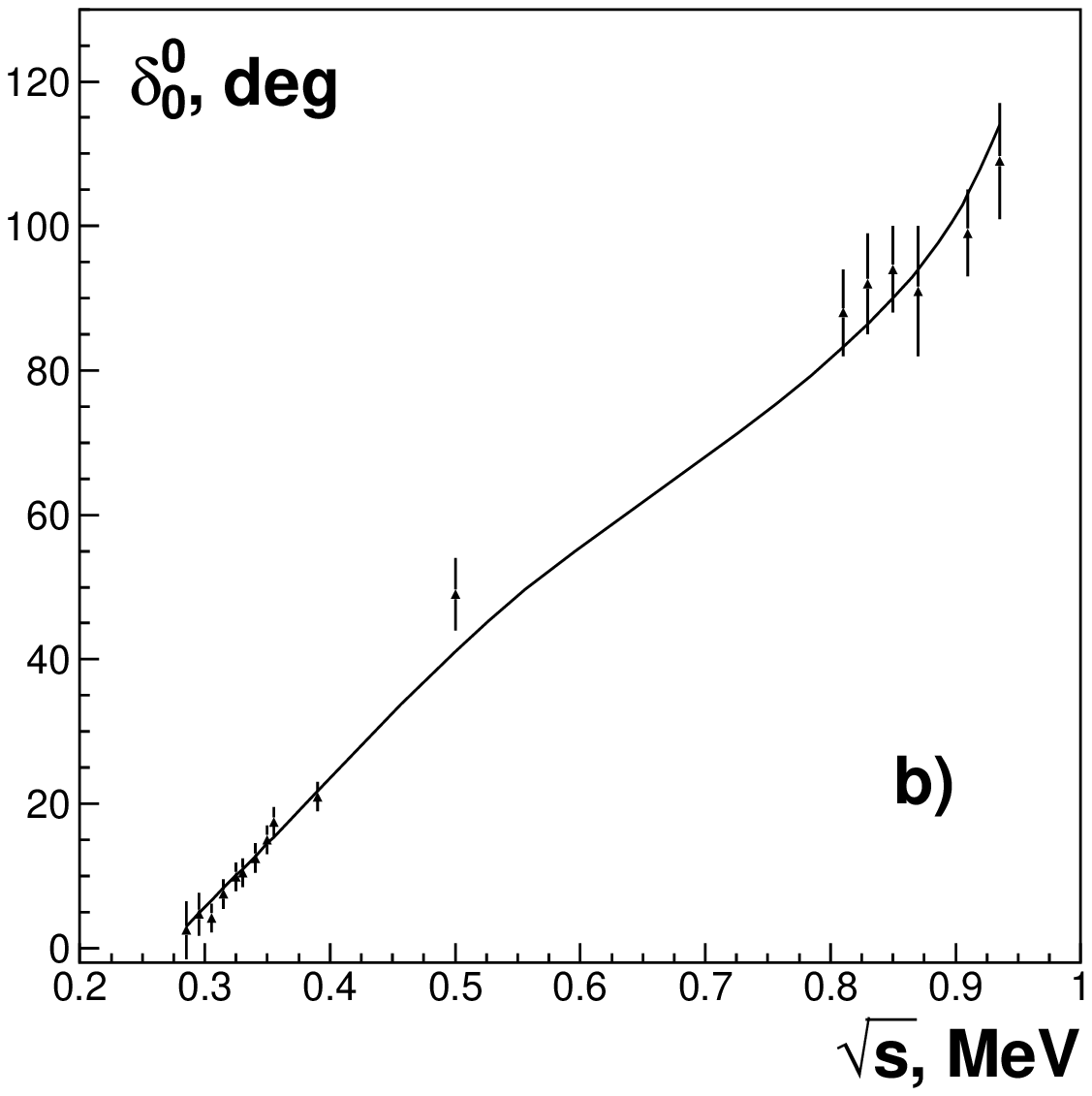,width=6cm}\\
\epsfig{file=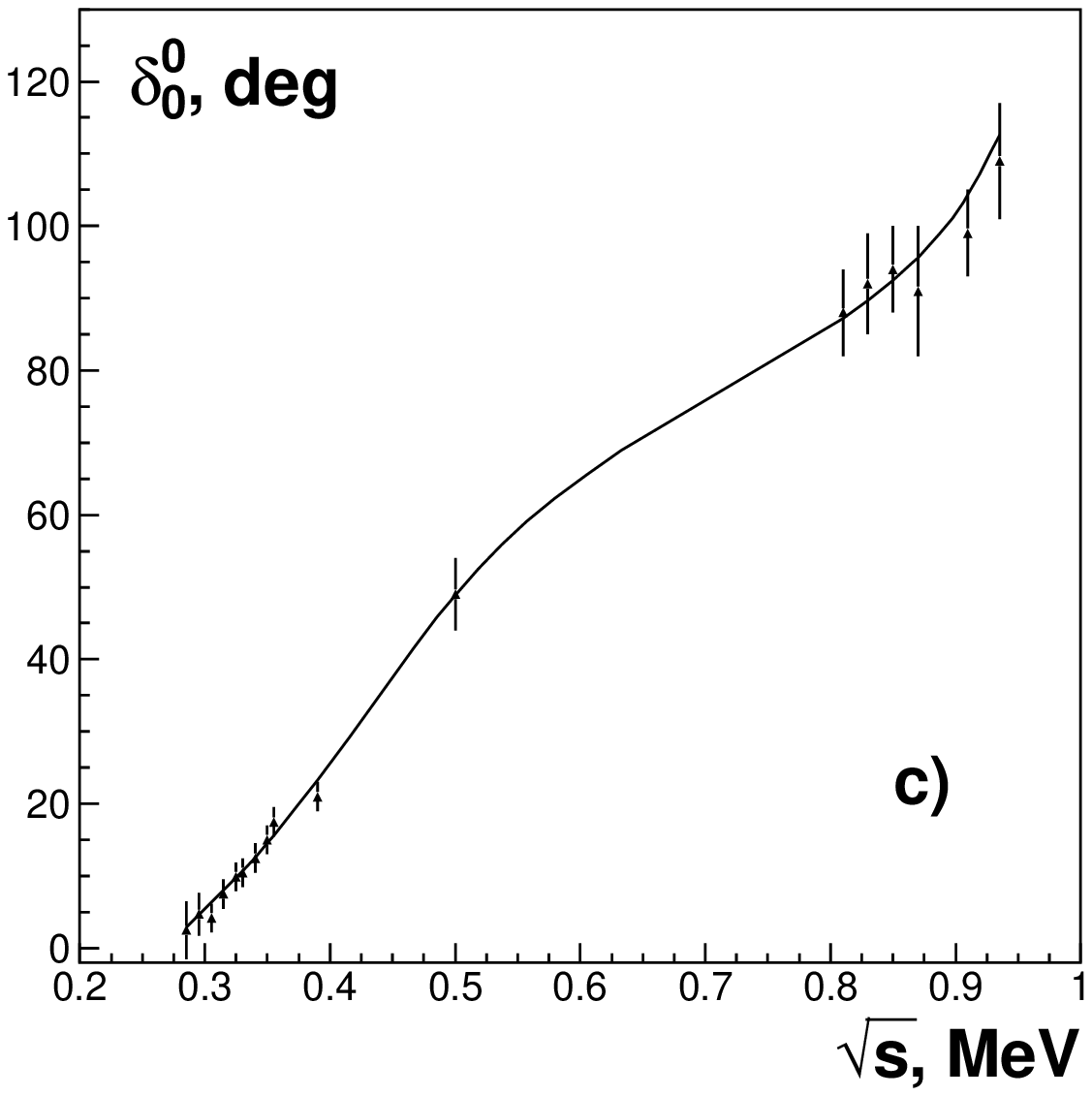,width=6cm}
\epsfig{file=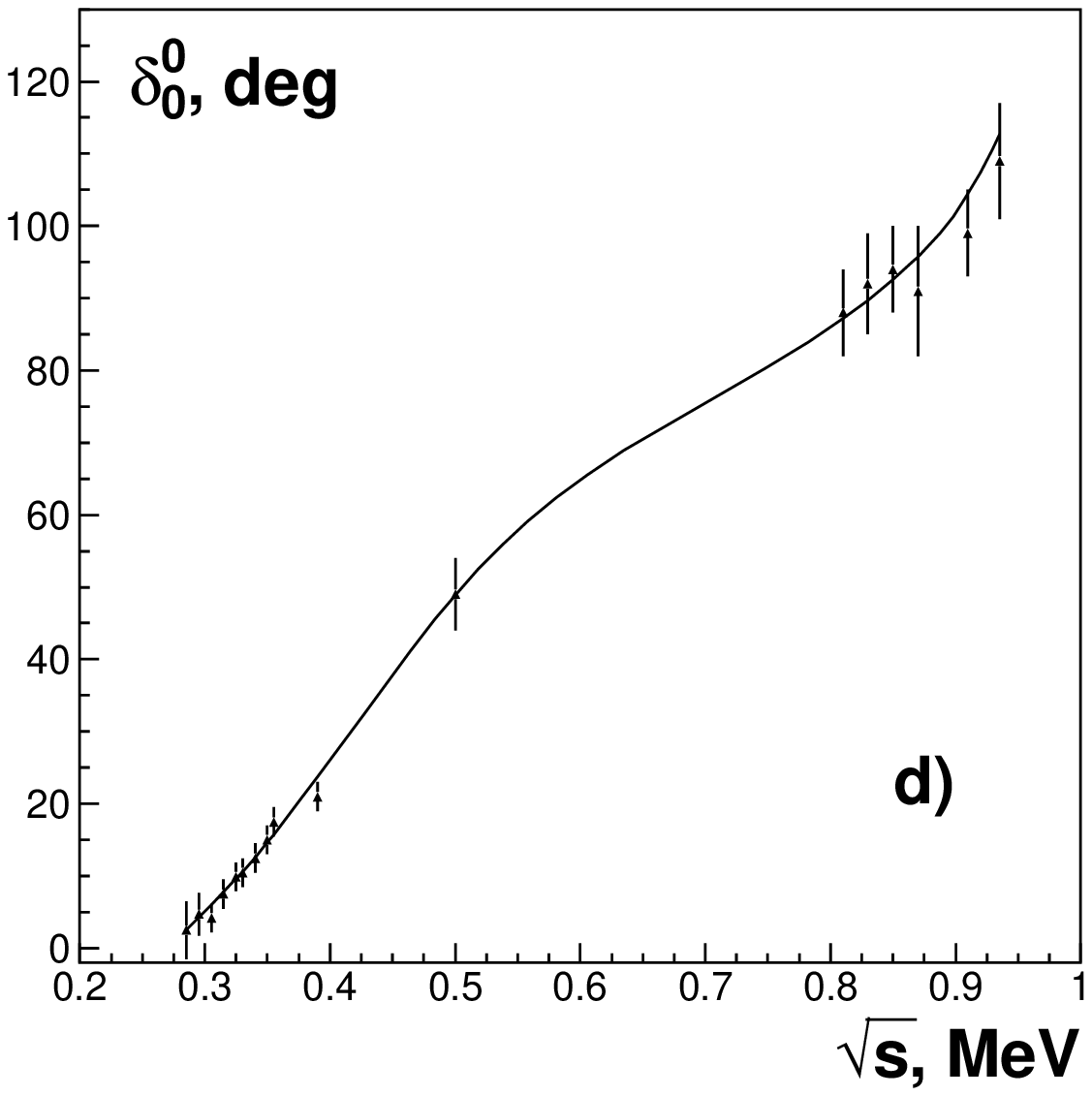,width=6cm}
 \caption{Description of the
$K_{e4}$ data with the $D$-matrix solutions 2 ,3 (with standard
errors for the point $\delta^0_0(500MeV)$ \protect\cite{Aloisio})
and solutions 4,5 (with the decreased error for this point).}
\label{delta}
\end{center} \end{figure}

D-matrix parameters can be expressed in the same terms (bare masses
and couplings) as parameters of a K-matrix fit. The subtraction
point for calculation of the real part of the loop diagrams, i.e.
$B^j_{\alpha\beta}(M_s^2)$ in eq. (\ref{D9}), was taken at the
corresponding two-particle threshold; the parameters $b^j$ were
optimized in the fit. In such an approach our data base can be
described with a very similar quality as in the framework of the
K-matrix approach, see Table~\ref{table_chi}. As expected, the
D-matrix fit provides a better description of the $K_{e4}$ data due
to the more correct behaviour of the amplitude near left-hand side
singularities. The behaviour of the phase shift $\delta^0_0$ and its
description in the mass region from the threshold to 1 GeV is shown
in Fig~\ref{delta}.

Below we present four D-matrix solutions: the bare masses and their
couplings are given in Table~\ref{table_const} (Solutions 2,3,4,5).
In Solutions 2,3 the $K_{e4}$ point near $500$ MeV was taken with
the error given by the experimental group. However, these solutions
do not reproduce this point satisfactory. To force the $\pi\pi$
phase shift to describe this point, we decreased the error by a
factor 10 and repeated the $D$-matrix fit of the data. In such an
approach we were able to describe the data at $500$ MeV rather well
(Solutions 4,5); however, we obtained a systematically worse
description of the proton-antiproton annihilation into the
$\pi^0\pi^0\pi^0$ and $\eta\eta\pi^0$ channels (see Table
~\ref{table_chi}).

The solution with the $1/s^2$ term included (Solutions 3,5) produced
a better total $\chi^2$ and a slightly better description of the
$K_{e4}$ data. The term $1/s^2$ can produce two additional poles in
the mass region below the $\pi\pi$ threshold. The pole in the mass
region around 400 MeV has moved to lower masses by about 80 MeV
compared to solutions without the $1/s^2$ term, see
(\ref{table_pole}), while the poles situated above 900 MeV
practically do not change their positions.

It is seen from Table~\ref{table_const} that the masses of bare
states are hardly changed from the K-matrix solution and most of the
couplings are shifted by less
 than 20\%. The positions of the amplitude poles above 900 MeV also
changed very little:
\be
\begin{tabular}{c|c|c|c|c|c}
             ~& Solution 1 & Solution 2 & Solution 3 & Solution 4 & Solution 5 \\
\hline
$\sigma$-meson& 420-i\,395 & 407-i\,281 & 365-i\,283 & 414-i\,186 & 406-i\,192 \\
 $f_0(980)$   &1014-i\, 31 &1015-i\, 36 &1012-i\, 31 &1005-i\, 20 &1005-i\, 23 \\
 $f_0(1300)$  &1302-i\,180 &1307-i\,137 &1303-i\,140 &1332-i\,140 &1326-i\,137 \\
 $f_0(1500)$  &1487-i\, 58 &1487-i\, 60 &1483-i\, 55 &1487-i\, 55 &1486-i\, 55 \\
 $f_0(1750)$  &1738-i\,152 &1781-i\,140 &1787-i\,143 &1795-i\,109 &1794-i\,114 \\
\end{tabular}\label{table_pole}
\ee
 The relative position of the poles and the
threshold singularity cuts is demonstrated in Fig. \ref{Xpolesco2}.

We see that a fit of the $K_{e4}$ data with the use of the
$D$-matrix approach unambiguously reveals the pole in the mass
region around 300-400 MeV, the low-mass $\sigma$-meson.

To trace the origin of the $\sigma$ pole, we multiplied all
couplings by the factor $\beta$ and the non-resonant terms by
$\beta^2$, and scanned this parameter from 1 (the physical
amplitude) to 0 (amplitude with poles corresponding to the bare
masses). Such an investigation shows that the $\sigma$-pole is
originated from the Adler regularization term. In the best fit the
Adler regularization point is optimized rather close to the physical
region $s_{A0}\simeq 0.15$ GeV$^2$. To check the stability of this
point we have performed the fit with this point fixed at
$s_{A0}=0.5$ $s_{A0}=1$ and $s_{A0}=1.5$ GeV$^2$. We observe a small
deterioration of the total $\chi^2$ due to a worse description of
the $\pi\pi\to K\bar K$ and $\pi\pi\to\eta\eta$ amplitudes. However,
the fit with $s_{A0}=0.5$ GeV$^2$ gives the best description for the
proton-antiproton annihilation into the $\pi^0\pi^0\pi^0$ channel
that is one of the most sensitive reactions to the description of
the low $\pi\pi$ mass region. The positions of the poles in all
three solutions coincide remarkably well and hence, we conclude that
the position of the $\sigma$-meson depends very little on the exact
position of the Adler regularization term.

\begin{table}[h]
\begin{center}
\caption\protect{List of the reactions and $\chi^2$ values for the
$K$-matrix and $D$-matrix solutions:
Solutions 3,5 with taken into account confinement interaction,
solutions 1,2,4 without it.}
\label{table_chi}
\begin{tabular}{|l|c|c|c|c|c|c|}
\hline
~ &Sol.     1&Sol.     2&Sol.     3&Sol.     4 &Sol.     5&N of\\
~       &K-matrix &D-matrix &D-matrix&D-matrix&D-matrix&  points \\
~           & 0 & 0 &$\sim 1/s^2$&0 &$\sim 1/s^2$&   \\
\hline
 ~& \multicolumn{6}{|c|}{The Crystal Barrel data} \\
\hline
from liquid $H_2$:              & ~     &~      &~    &~    &~     & ~   \\
$\bar pp\to \pi^0\pi^0\pi^0$    & 1.32  &1.37   &1.39 &1.44 &1.45  & 7110\\
$\bar pp\to \pi^0\eta\eta$      & 1.33  &1.34   &1.34 &1.33 &1.33  & 3595\\
$\bar pp\to \pi^0\pi^0\eta$     & 1.24  &1.33   &1.33 &1.55 &1.55  & 3475\\
\hline
from gaseous $H_2$:             &~      &      &      &     &     &  ~ \\
$\bar pp\to \pi^0\pi^0\pi^0$    & 1.39  &1.44  &1.45  &1.48 &1.49 &  4891\\
$\bar pp\to \pi^0\eta\eta$      & 1.31  &1.34  &1.30  &1.43 &1.31 &  1182\\
$\bar pp\to \pi^0\pi^0\eta$     & 1.20  &1.22  &1.22  &1.31 &1.32 &  3631\\
\hline
from liquid $H_2$:              & ~     &~     &~     &~     &~    &  ~ \\
$\bar pp\to \pi^+\pi^0\pi^-$    & 1.54  &1.46  & 1.45 & 1.46 & 1.47&  1334\\
from liquid $D_2$:              & ~     &~     & ~    & ~    & ~   &   ~ \\
$\bar pn\to \pi^0\pi^0\pi^-$    & 1.51  &1.47  & 1.47 & 1.46 & 1.46&  825 \\
$\bar pn\to \pi^-\pi^-\pi^+$    & 1.61  &1.54  & 1.55 & 1.50 & 1.51&  823 \\
\hline
from liquid $H_2$:              & ~     &~     &~     &~    &~    &  ~   \\
$\bar pp\to K_SK_S\pi^0$        & 1.09  &1.10  &1.10  &1.10 &1.10 &   394\\
$\bar pn\to K^+K^-\pi^0$        & 0.98  &1.00  &1.00  &1.03 &1.02 &   521 \\
$\bar pn\to K_LK^\pm\pi^\mp$    & 0.78  &0.79  &0.79  &0.79 &0.79 &   737 \\
from liquid $D_2$:              & ~     &~     &~     &~    &~    &  ~ \\
$\bar pp\to K_S K_S\pi^-$       & 1.66  &1.64  &1.64  &1.64 &1.63 &   396\\
$\bar pn\to K_S K^-\pi^0$       & 1.33  &1.31  &1.31  &1.31 &1.31 &   378 \\
\hline
~& \multicolumn{6}{|c|}{The GAMS data} \\
\hline
$\pi\pi\to(\pi^0\pi^0)_{S-wave}$& 1.23  &1.13  &1.15  &1.32 &1.30 & 68 \\
$\pi\pi\to(\eta\eta)_{S-wave}$  & 1.02  &1.05  &1.03  &1.58 &1.43 & 15 \\
$\pi\pi\to (\eta\eta')_{S-wave}$& 0.45  &0.30  &0.35  &0.35 &0.34 &  9  \\
\hline
~& \multicolumn{6}{|c|}{The BNL data} \\
\hline
$\pi\pi\to (K\bar K)_{S-wave}$  & 1.32  &1.13  &1.14  &0.97 &1.07 & 35 \\
\hline
 ~& \multicolumn{6}{|c|}{The CERN-Munich data: $Y^0_0\ \ldots\ Y^1_6$ } \\
\hline
$\pi^-\pi^+\to\pi^-\pi^+$
                                & 1.82  &1.86  &1.86  &2.05 &2.03 & 705\\
 \hline
 ~& \multicolumn{6}{|c|}{The $K_{e4}$ decay data} \\
\hline
$\delta^0_0(\pi^-\pi^+\to\pi^-\pi^+)$
                                & 1.51  &1.02  &0.84  &0.80 &0.83 & 17 \\
\hline
\end{tabular}
\end{center}
\end{table}

\begin{table}[ht]
\begin{center}
\caption\protect{The $f_0^{\rm bare}$-resonances: masses $M_n$ (in MeV
units), decay coupling constants $g_n$ of Table 1 (in GeV units),
mixing angles (in degrees), background terms $f_n$ and confinement
singularity term $G/s^2$ (factor $G$ in GeV units). }\\
 ~\\
\label{table_const}
\begin{tabular}{|l|c|c|c|c|c|}
\hline
~ &Solution 1&Solution 2&Solution 3&Solution 4&Solution 5 \\
\hline
$M_1$ & 671  &  685  & 697 & 611 & 615 \\
$M_2$ & 1205 &  1135 & 1135& 1078& 1096\\
$M_3$ & 1560 &  1561 & 1558& 1575& 1572\\
$M_4$ & 1210 &  1290 & 1284& 1334& 1330\\
$M_5$ & 1816 &  1850 & 1848& 1858& 1857\\
\hline
$g_{1}$ & 0.860 &  0.926 & 0.892 & 1.090& 1.083\\
$g_{2}$ & 0.956 &  0.950 & 0.935 & 0.099& 1.066\\
$g_{3}$ & 0.373 &  0.290 & 0.284 & 0.302& 0.302\\
$g_{4}$ & 0.447 &  0.307 & 0.308 & 0.264& 0.275\\
$g_{5}$ & 0.458 &  0.369 & 0.370 & 0.317& 0.330\\
\hline
$g^{(1)}_{\eta\eta}$ &-0.382 & -0.213 &-0.232 &-0.176 &-0.193 \\
$g^{(1)}_{\eta\eta'}$&-0.322 & -0.500 &-0.500 &-0.500 &-0.500 \\
\hline
$g^{(1)}_{4\pi}$, $g^{(2)}_{4\pi}$   & 0     &  0     & 0     & 0     & 0 \\
$g^{(3)}_{4\pi}$     & 0.638 &  0.534 & 0.530 & 0.511 & 0.514 \\
$g^{(4)}_{4\pi}$     & 0.997 &  0.790 & 0.794 & 0.691 & 0.702 \\
$g^{(5)}_{4\pi}$     & -0.901&  -0.862& -0.856& -0.797& -0.814\\
\hline
$\Phi_{1}$ & -74 &  -83 & -82& -81& -82 \\
$\Phi_{2}$ & 6   &  -2.6& -1.9& -1.1& -2.4\\
$\Phi_{3}$ & 9   &  5   & 5  & 5  & 5   \\
$\Phi_{4}$ & 38  &  31  & 32 & 25 & 25  \\
$\Phi_{5}$ & -64 &  -71 & -68& -77& -77 \\
\hline
$f_{\pi\pi\to \pi\pi}   $& 0.337  &  0.408  & 0.358 & 0.763 & 0.687 \\
$f_{\pi\pi\to K\bar K}  $& 0.212  &  0.036  & 0.044 & 0.103 & 0.065 \\
$f_{\pi\pi\to 4\pi}     $& -0.199 &  -0.101 & -0.092& -0.051& -0.062   \\
$f_{\pi\pi\to \eta\eta} $& 0.389  &  0.438  & 0.413 & 0.538 & 0.512    \\
$f_{\pi\pi\to \eta\eta'}$& 0.394  &  0.518  & 0.485 & 0.610 & 0.597    \\
\hline
$G/s^2$& 0  &  0 &$ -0.00077/s^2$ & 0 &$-0.00071/s^2$ \\ \hline
\end{tabular}
\end{center}
\end{table}

\subsection{Calculation of the scattering length}

In our expression for the $\pi\pi$ scattering amplitude which takes
into account the $\pi^0\pi^0$ and $\pi^+\pi^-$ phase volumes the
$\pi\pi$-phase does not go to zero on the threshold of two charged
pions, see Appendix C. We calculate the scattering length of the
$\pi^+\pi^-$ system at the threshold of two charged pions using the
following expression:
 \be
\label{scat_len}
&&\frac{s^{1/2}}{m_{\pi^+}\!+\!m_{\pi^-}}{\rm Re } \left
[\sin\delta_0^{(0)}e^{i\delta_0^{(0)} }\right ]_{k\to 0}\simeq
c_0^{(\pm)}+a_0^{(\pm)}k+b_0^{(\pm)} k^3\, ,
\nn \\
&&{\rm with}\quad a_0=\frac 32 a_0^{(\pm)},\qquad k=\frac 12
\sqrt{s-(m_{\pi^+}\!+\!m_{\pi^-})^2}\quad {\rm and}\quad
m_{\pi^+}=m_{\pi^-}.
\ee
 The scattering length values extracted from the $D$-matrix
solutions are equal to:
\be \label{scat_len_v}
 \begin{tabular}{c|c|c|c}
 Solution 2 &  Solution 3&  Solution 4& Solution 5 \\
\hline
  0.253$m_{\pi}^{-1}$&0.209$m_{\pi}^{-1}$&0.204$m_{\pi}^{-1}$&
0.177$m_{\pi}^{-1}$\\
\end{tabular}
\ee
 It is seen that the inclusion of the $1/s^2$ term decreases the
scattering length by $\sim 0.05$ $m_\pi^{-1}$ and a similar effect
 comes from a precise description of the $K_{e4}$ experimental
 point at $500$ MeV.

The amplitude phase was extracted by the $K_{e4}$ collaboration
under the assumption that it is equal to zero at the threshold of
two charged pions. Then there is a question about the uncertainty
which appears when these data are fitted with an expression which
takes into account exactly the thresholds of neutral and charged
pions. To check this we put in Solution 2 all pion masses equal to
the mass of a charged pion. As expected, notable deteriorations were
observed only in the proton-antiproton annihilation into three
neutral pions and at low energy points for the $K_{e4}$ data. With a
very small tuning of the parameters we obtained very similar
$\chi^2$ values for the description of the $K_{e4}$ data. The
scattering length which in this case is calculated as $a_0=\frac 32
a_0^{(\pm)}$ appeared to be 0.248 $m_\pi^{-1}$. Then, with this
parameters fixed, we introduced back the difference between neutral
and charged pion thresholds but not refitted the data. The
scattering length obtained by eq. (\ref{scat_len}) was found to be
0.260 $m_\pi^{-1}$: a value which is very close to that obtained in
Solution 2. Thus, we conclude that the investigated uncertainty is
less than 0.010 $m_\pi^{-1}$ and is smaller than the systematic
error which comes from different parameterizations of the amplitude.

It is instructive to compare the results of eq. (\ref{scat_len_v})
with those obtained without taking into account different values of
$\pi^0\pi^0$ and $\pi^+\pi^-$ threshold singularities:
$a_0=(0.233\pm 0.013) \mu^{-1}_\pi$\cite{Garcia}, $a_0=(0.220\pm
0.005) \mu^{-1}_\pi$\cite{Gasser}.

\section{Conclusion}

The analysis of the large data sets performed in the framework of
the $K$-matrix and $D$-matrix approaches demonstrates a very good
stability for the amplitude parameters and pole positions above 900
MeV. The description of the $K_{e4}$ data demands the presence of
the pole slightly above the $\pi\pi$ threshold.  The pole position
was found to be at $390\pm 45-i\,210\pm 50$ MeV and the scattering
length $0.215\pm 0.040$ $m_\pi^{-1}$.

The confinement singularity, $1/s^2$, slightly improves the overall
description but is not crucial for a good description of the
$K_{e4}$ data and for the existence of the $\sigma$ meson pole
singularity. However, the presence of such a term influences the
pole position shifting it to lower masses by about 100 MeV and
shifting the scattering length to lower values (by $\sim 0.05$
$m_\pi^{-1}$).

The imaginary part of the $\sigma$ pole position in the solutions
which fit precisely the data point at $500$ MeV is lower by about
100 MeV compare to the solutions where the fit is not forced to
describe this point. The scattering length in such solutions is also
systematically shifted to lower values by $\sim 0.05$
$\mu_\pi^{-1}$.

Within the description of the $00^{++}$-wave in the channels
$\pi\pi$, $\pi\pi\pi\pi$, $K\bar K$, $\eta\eta'$ we obtain the
following complex masses of the $f_0$ resonances:
 \bea
 f_0(980)&\qquad & M_I=1011\!\pm\! 5- i\,31\!\pm\! 4\,\, {\rm MeV} \nn\\
         &\qquad & M_{II}=1035\!\pm\!50- i\,460\!\pm\! 50\,\, {\rm MeV} \nn\\
 f_0(1300)&\qquad & M=1285\!\pm\!30- i\,160\!\pm\!20\,\, {\rm MeV} \nn\\
 f_0(1500)&\qquad & M=~1488\!\pm\! 4\,- i\,53\!\pm\! 5\,\, {\rm MeV} \nn\\
 f_0(1790)&\qquad & M=1775\!\pm\! 25- i\,140\!\pm\! 15\,\, {\rm MeV}
 \label{1a}
 \eea
The masses of the $D$-matrix approach, eq. (\ref{1a}), coincide well
with those obtained in the $K$-matrix approximation \cite{book3}.
The $f_0(980)$ is determined by two poles, on the second (under the
$\pi\pi$ threshold) and third (under the $\pi\pi$ and $K\bar K$
thresholds) sheets -- the same splitting of poles we have in the
$K$-matrix solutions \cite{ANS-content}.

For the low mass region the solution with $1/s^2$ singularity gives
several poles on the second sheet:
 \bea
 f_0(\sigma_{I})\qquad M=365\!\pm\! 15- i\,283\!\pm\! 12\,\, {\rm MeV} \nn\\
 f_0(\sigma_{II})\qquad M=80\!\pm\! 10- i\,187\!\pm\! 15\,\, {\rm MeV} \nn\\
 f_0(\sigma_{III})\qquad M=-94\!\pm\! 12- i\,93\!\pm\! 10\,\, {\rm MeV}\,
 .
 \label{2}
 \eea
If the fit is forced to describe the $K_{e4}$ experimental point at
$500$ MeV, we have:
 \bea
 f_0(\sigma_{I})\qquad M=406\!\pm\! 15- i\,192\!\pm\! 15\,\, {\rm MeV} \nn\\
 f_0(\sigma_{II})\qquad M=74\!\pm\! 10- i\,190\!\pm\! 50\,\, {\rm MeV} \nn\\
 f_0(\sigma_{III})\qquad M=-96\!\pm\! 22- i\,100\!\pm\! 25\,\,{\rm MeV}
 \, .
 \label{2a}
 \eea
We also test the changes in the description of data with an
elimination of the $1/s^{2}$ singularity. In this case the fit to
the data gives the masses of the $f_0$ resonances at $\sqrt s > 900$
MeV practically the same as in ref. (\ref{1a}) -- the changes are in
the low-mass pole structure. Without the $1/s^2$ singularity, the
position of the $\sigma$ pole in the fit, neglecting the $500$ MeV
point, gives:
 \bea
 f_0(\sigma_{I})\qquad M=407\!\pm\! 12- i\,289\!\pm\! 10\,\, {\rm MeV} \nn\\
 \label{2b}
 \eea
and with the fit forced to describe the $500$ MeV point:
 \bea
 f_0(\sigma_{I})\qquad M=412\!\pm\! 12- i\,186\!\pm\! 15\,\, {\rm MeV}
 \label{2c}
 \eea
 So, the $\sigma$-meson arises as a pole near the $\pi\pi$ threshold in
 both versions, with and without including the confinement singularity
($1/s^2$) into the $\pi\pi$ scattering block. Though the confinement
singularity leads to the appearance of several poles under the
$\pi\pi$ cut, it is hardly possible to distinguish these two
versions on the basis of the data.

 \centerline{\bf Acknowledgement}

The paper was partially supported by the grant RSGSS-3628.2008.2.

\section {Appendix A: Examples of Sets of Diagrams Resulting in
$1/t^2$ Singularities}

\begin{figure}
\centerline{\epsfig{file=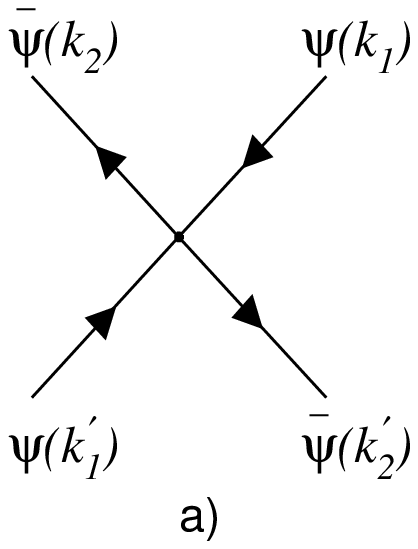,width=4cm}
            \epsfig{file=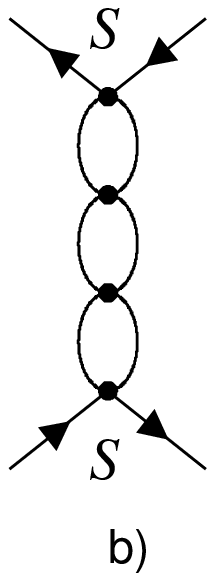,width=4cm}\hspace{-1.0cm}
            \epsfig{file=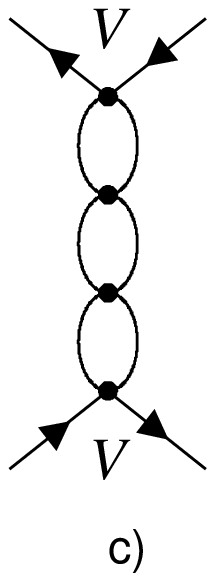,width=4cm}}
\caption{Interaction block (a) and sets of loop diagrams for
S and V exchanges.
\label{B1}}
\end{figure}

Here we consider, as an example, the confinement set of the loop
diagrams, Fig. \ref{X2f22}c, and present an illustrative calculation
which results in singularities of the $1/t^2$ type in scalar and
vector channels. We use the following interaction blocks, see Fig.
\ref{B1}a:
\bea
&&\sum\limits_n \Psi_{meson(n)}\Psi^*_{meson(n)}\to
G_S^{(L)}\bigg((k_1-k_2)^2\bigg)\psi(k_1)\bar \psi(k_2)
G_S^{(R)}\bigg((k'_1-k'_2)^2\bigg)
\psi(k'_1)\bar \psi(k'_2) \nn\\
&&
+G_V^{(L)}\bigg((k_1-k_2)^2\bigg)\psi(k_1)\gamma_\mu\bar \psi(k_2)
G_V^{(R)}\bigg((k'_1-k'_2)^2\bigg)
\psi(k'_1)\gamma_\mu\bar \psi(k'_2)\, .
\eea
 Then the confinement interaction turns into a set of the loop diagrams,
Fig. \ref{B1}b and Fig. \ref{B1}c. The scalar and vector exchanges,
correspondingly, read:
\bea
V_S(t)=\frac{B_S(t)}{1-B_S(t)}\, ,\quad
V_{V;\mu\nu}(t)=-\delta^\perp_{\mu\nu}\frac{B_V(t)}{1-B_V(t)}\, .
\eea
 For a scalar loop diagram one has:
\bea
&&B_S(t)=\int\limits_{4m^2}^\infty\frac{dt'}{\pi}
d\Phi_2 (P';k'_1, -k'_2)\frac{N_S(t')\, Sp [(\hat k'_2-m)
(\hat k'_1+m)]}{t'-t-i0}\, ,\nn\\
&&d\Phi_2 (P';k'_1, -k'_2)=\frac{1}{2(2\pi)^2}d^4k'_1d^4k'_2
\delta(k'^2_1-m^2)\delta(k'^2_2-m^2) \to
\frac{1}{16\pi}\sqrt{1-\frac{4m^2}{t'}}\,,\nn\\ && Sp [(\hat
k'_2-m)(\hat k'_1+m)]\to 2(t'-4m^2)\, .
\eea

Here we replace $ G_S^{(R)}G_S^{(L)}\to N_S$.
 An analogous loop for vector exchange reads:
\bea
&&-\delta^\perp_{\mu\nu} B_V(t)=
\int\limits_{4m^2}^\infty\frac{dt'}{\pi} d\Phi_2 (P';k'_1,
-k'_2)\frac{N_V(t')\, Sp [\gamma^\perp_\mu(\hat
k'_2-m)\gamma^\perp_\nu (\hat k'_1+m)]}{t'-t-i0}\, \nn\\
&&=-\delta^\perp_{\mu\nu}
\int\limits_{4m^2}^\infty\frac{dt'}{\pi}
\frac{G_V^2(t')(2m^2+t')}{t'-t-i0}\,
\frac{1}{16\pi}\sqrt{1-\frac{4m^2}{t'}}\,
 . \eea
Confinement singularities appear if $B_S(t)$ and $B_V(t)$ behave
near $t=0$ as follows:
\be
B_S(t)=1-\frac{t^2}{\beta_S}+O(t^3),\qquad
 B_V(t)=1-\frac{t^2}{\beta_V}+O(t^3)\, ,
\ee
that means the requirements
\be
\frac{d}{dt}B_S(t)\bigg|_{t=0}=0\, , \qquad
\frac{d}{dt}B_V(t)\bigg|_{t=0}=0 \, .
\ee

\section{Appendix B: Simplified Consideration of\\ the
$00^{++}$ Wave in the Low-Energy Region}

The partial pion--pion scattering amplitude being
 a function of the invariant energy squared,
$s=M^2$, can be represented as a ratio $N(s)/D(s)$
\cite{WMandelstam}), where $N(s)$ has a left-hand cut due to the
``forces'' (the interactions caused by the $t$- and $u$-channel
exchanges), and the function $D(s)$ is determined by the
rescattering in the $s$-channel.  The standard presentation of the
N/D-method may be found, for example, in \cite{WChew}).

The $\pi\pi$ scattering block related to the $1/s^2$ singularity
reads:
\be \label{ND1}
G(s)\frac {1}{s^2} G(s).
\ee
 The $s$-channel re-scatterings give a set of divergent terms which
convolutes into the following unitary amplitude:
\bea \label{ND2}
 &&A(s)=G(s)\frac {1}{s^2} G(s)+G(s)\frac
{1}{s^2}\Pi(s)\frac {1}{s^2} G(s)+...\nn\\
&&=\frac{G^2(s)}{s^2-\Pi (s)}= G^2(s)\bigg
[s^2-\int\limits_{4\mu^2_{\pi}}^{\infty}\frac{ds'}{\pi} \frac{G^2
(s')\rho(s')}{s'-s}\bigg ]^{-1}
\eea
 Here $\rho(s)$ is the invariant $\pi\pi$ phase space.
In the physical region, at $s>4m^2_{\pi}$ and $s$ on the upper edge
of the threshold cut, we have:
 \be \label{ND4}
\Pi (s)= \int\limits_{4\mu^2_{\pi}}^{\infty}\frac{ds'}{\pi}
\frac{G^2 (s')\rho(s')}{s'-s-i0}=
P\int\limits_{4\mu^2_{\pi}}^{\infty}\frac{ds'}{\pi} \frac{G^2
(s')\rho(s')}{s'-s}+iG^2 (s)\rho(s)
\ee
with the following relation to the $IJ^{PC}=00^{++}$ phase shift:
 $\rho(s)A(s)=\exp{\bigg(i\delta^0_0(s)\bigg)}\sin \delta^0_0(s)$
The product of the vertices $G^2 (s)$ is actually an $N$-function,
and we re-write $G^2 (s)\to N(s)$; this allows to present the
amplitude (\ref{ND2}) as
\be
 A(s)=\frac{N(s)}{D(s)}\; ,
\;\;\;D(s)=s^2- \int \limits_{4m^2_\pi}^\infty \frac {ds'}{\pi}
\frac{\rho(s')N(s')}{s'-s-i0}\; .\nn
\label{ND6}
\ee
 The $N$-function, being determined by the left-hand singularities
caused by forces due to $t$-channel and $u$-channel meson exchanges,
is written as an integral along the left cut as follows:
\be
\label{ND7}
 N(s)=\int \limits_{-\infty}^{s_L}
\frac{ds'}{\pi}\frac{L(s')}{s'-s}\; ,
\ee
 where the value $s_L$ marks the beginning of the left-hand cut.
For example, for the one-meson exchange diagram $g^2/(m^2 -t)$ the
left-hand cut starts at $s_L=4m_\pi^2-m^2$, and the $N$-function in
this point has a logarithmic singularity; for the two-pion exchange,
$s_L=0$.

We replace the left-hand integral for $N(s)$, eq.(\ref{ND7}), by the
following sum:
 \be \label{ND12}
 N(s)=\int\limits_{-\infty}^{s_L} \frac{ds'}{\pi}\frac{L(s')}{s'-s}\to
16\pi\sqrt s\,\sum\limits_n \frac {L_n}{s_n-s}\; ,
\ee
 where $L_n$ and $s_n$ are ``force parameters'', $-\infty <s_n< s_L$.

The pole approximation {\it ansatz} (\ref{ND12}) allows us calculate
the scattering amplitude in the physical region:
\be \label{ND13}
\exp{\bigg(i\delta^0_0(s)\bigg)}\sin \delta^0_0(s)
=\frac{\sqrt{s-4m^2_\pi}\sum\limits_n  L_n(s-s_n)^{-1}}{s^2-
\sum\limits_n\bigg( \sqrt{4m^2_\pi-s_n} +i\sqrt{s-4m^2_\pi}\bigg)
L_n(s-s_n)^{-1}}\,.
\ee

\begin{figure}[h]
\centerline{\epsfig{file=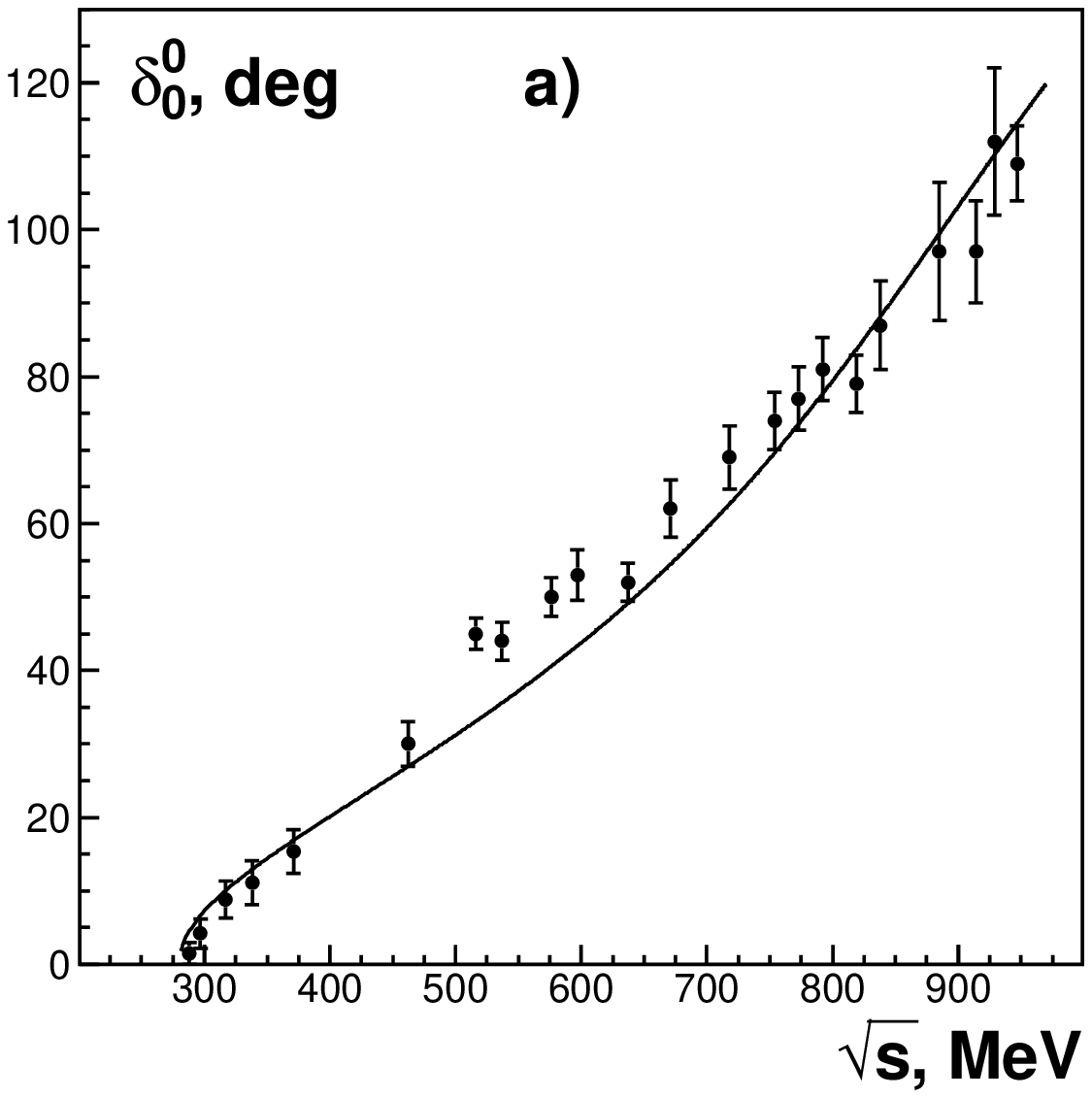,width=6cm}
            \epsfig{file=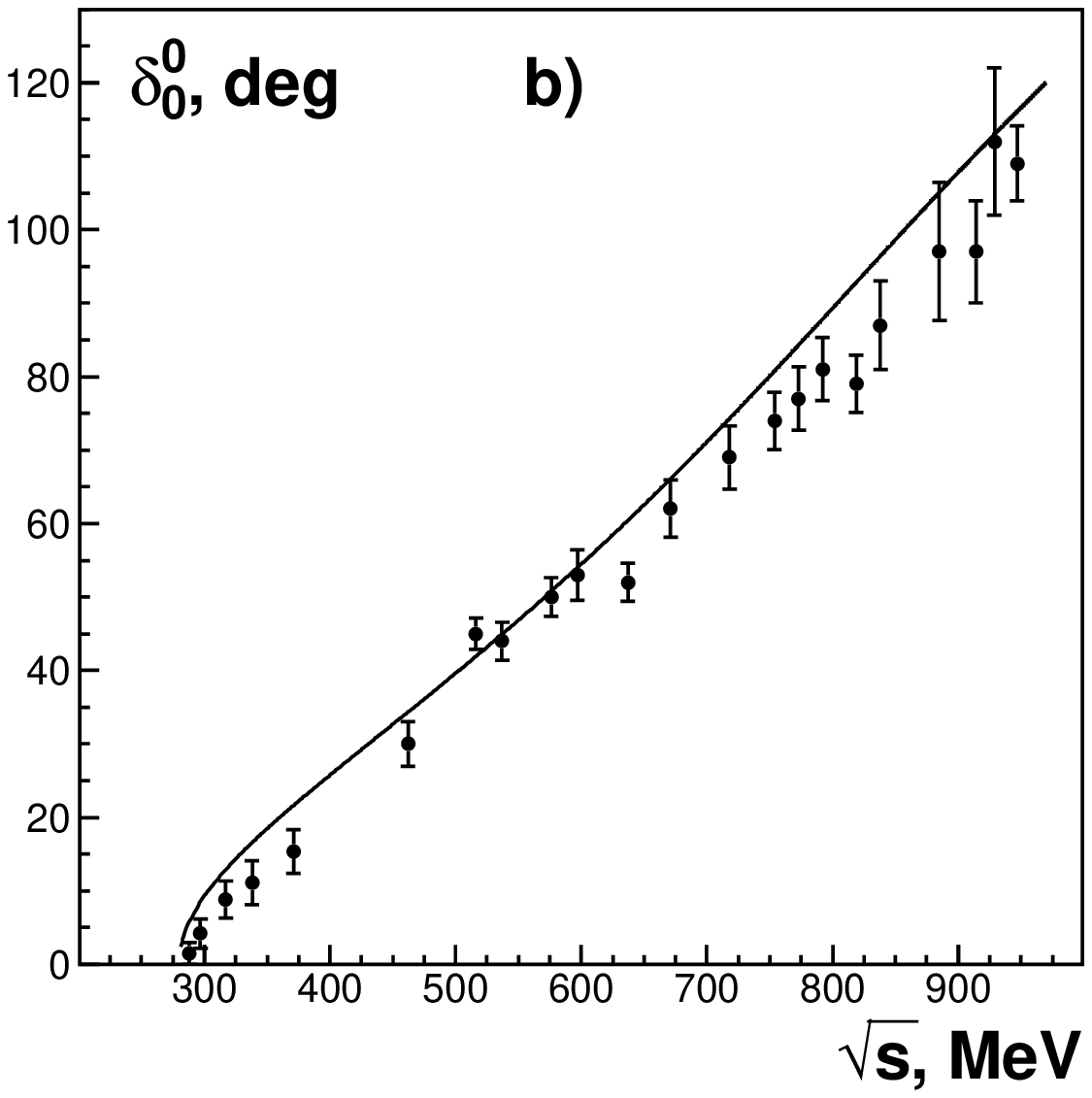,width=6cm}}
\centerline{\epsfig{file=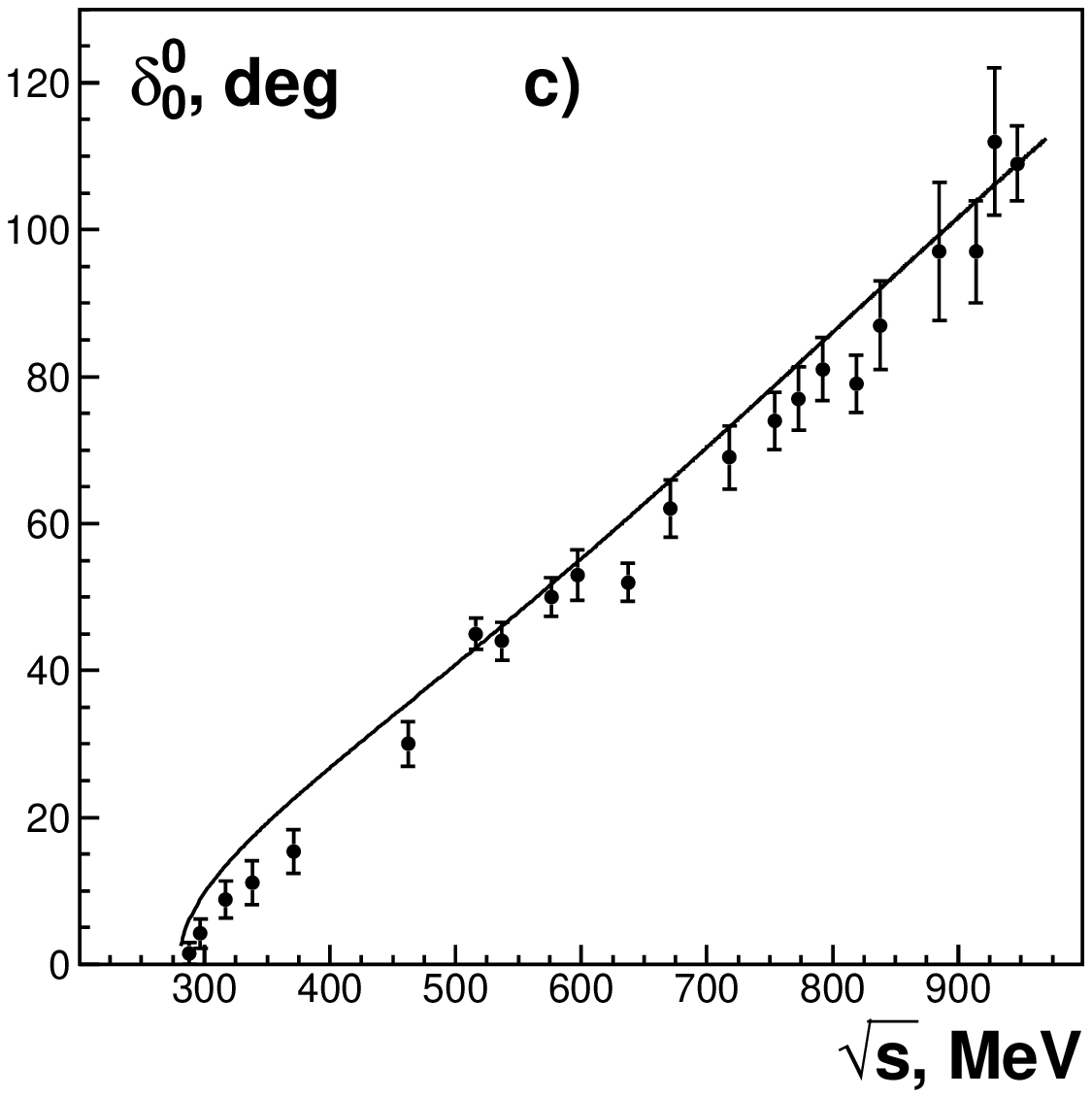,width=6cm}
            \epsfig{file=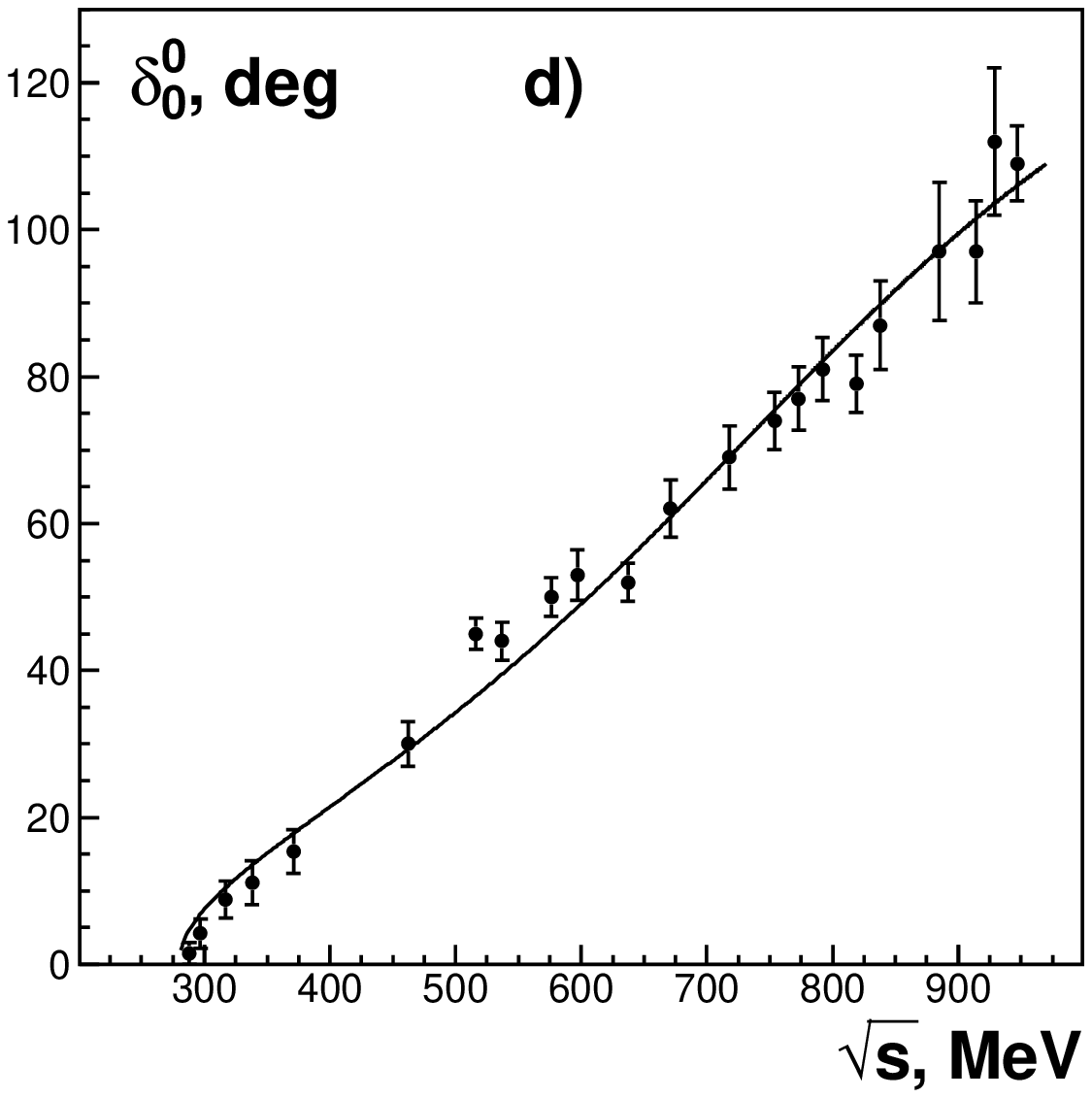,width=6cm}}
\caption{Examples of the fit of low energy data
\protect\cite{km,Pislak} in terms of eq. (\protect\ref{nc6})
\label{nc-f1}
}
\end{figure}

Here we give an example of a very simple, and formally correct,
consideration of the $00^{++}$ wave in the low-energy region. Using
eq. (\ref{ND13}) we write an analytic and unitary amplitude as
follows:
\beq \label{nc5}
 A^{I}_{thr}(s) = \frac{g^2}{s^2-\bigg(a_{I} +b_{I}s+ ig^2
\sqrt{s-4\mu^2_\pi}\bigg)}\, ,
\eeq
and hence,
\bea \label{nc6}
 && \exp[2i\delta^0_0(s)]=\frac{D^{I}(s)}{D^{I*}(s)}=
\frac{[k-(a+ib)][k-(-a+ib)][k-(c+id)][k-(-c+id)]}{[k-(a-ib)]
[k-(-a-ib)][k-(c-id)][k-(-c-id)]}\nn \\
&& {\rm with}\quad a>0,\quad b>0\, .
\eea
Considering $(a,b,c,b)$ as parameters, we fit the data for
$\delta^0_0(s)$ in the energy interval $280\leq\sqrt{s}\leq 950$
MeV, see Fig. \ref{nc-f1}. We obtain the following parameters and
amplitude pole positions, $M_{I}$ and $M_{II}$:
 \bea {\rm \,Fig.4a}:
&&\quad a=3.1\mu_\pi,\quad  b=1.0\mu_\pi,
          \quad  c=7.7\mu_\pi,\quad d=9.0\mu_\pi, \nn\\
          &&M_I=(896.3   -i274.3){\rm MeV},\quad
            M_{II}= (2163.7 -i2511.0){\rm MeV}     \nn\\
{\rm \,Fig.4b}:&&\quad a=2.8\mu_\pi,\quad , b=1.5\mu_\pi,\quad ,
          c=7.7\mu_\pi,\quad   d=4.5\mu_\pi,\quad \nn\\
          &&M_I=(828.8   -i391.4){\rm MeV},\quad
            M_{II}= (2166.6 -i1252.1){\rm MeV} \nn\\
{\rm \,Fig.4c}:&&\quad a=2.6\mu_\pi,\quad  b=1.9\mu_\pi,\quad
           c=5.4\mu_\pi,\quad   d=1.0\mu_\pi,\quad    \nn\\
          &&M_I=(759.7   -i509.1){\rm MeV},\quad
            M_{II}= (1529.8 -i275.4){\rm MeV} \nn\\
{\rm \,Fig.4d}:&&\quad a=2.5\mu_\pi,\quad  b=1.3\mu_\pi,\quad
           c=50.0\mu_\pi,\quad  d=2.0\mu_\pi,\quad  \nn\\
          &&M_I=(742.1   -i352.2){\rm MeV},\quad
            M_{II}= (14002.8 -i559.9){\rm MeV}\, .
\label{ND17}
\eea
In all solutions the scalar-isoscalar scattering length  is not
small: $a^0_0\sim (0.3-0.4)\mu_{\pi}^{-1}$ .

We use in the fit of Fig. \ref{nc-f1} the values for $\delta^0_0$
found in \cite{km} in order to perform a more visual comparison of
the obtained here results, eq. (\ref{ND17}), with those in
\cite{AN}. Let us recall that we fit in \cite{AN} the amplitude
$00^{++}$ in the region $280\leq\sqrt{s}\leq 900$ MeV in the
framework of the dispersion relation approach sewing the
$N/D$-solution with the K-matrix one at $450\leq\sqrt{s}\leq 1950$
MeV. Taking into account the left-hand cut contribution (it was a
fitting function), we obtained in \cite{AN} the best fit with the
$\sigma$-meson pole at $M_{\sigma}=(430\pm 150) -i(320\pm 130)$ MeV.
So, the accounting for the left-hand cut and data at $\sqrt{s}> 900$
MeV results in a smaller value of the $M_{\sigma}$.

In the approaches, which take into account the left-hand cut as a
contribution of some known meson exchanges, the pole positions  were
obtained at low masses as well. For example, the dispersion relation
approach results: $M_\sigma \simeq (470-i460)$ MeV
\cite{WBasdevant}, $M_\sigma \simeq (450-i375)$ MeV \cite{WZinn},
and the meson exchange models give:
 $M_\sigma \simeq (460-i450)$ MeV \cite{WBugg},
 $M_\sigma \simeq (400-i60)$ MeV \cite{WSpeth}.

\section {Appendix C: The $\pi\pi$ Scattering Amplitude near Two-Pion
Thresholds}

Here we consider the $\pi\pi$ scattering amplitude near two-pion
thresholds taking into account the mass difference of charged and
neutral pion systems, $\pi^+\pi^-$ and $\pi^0\pi^0$.

The following $\pi\pi$-amplitudes describe scattering reactions near
the thresholds:
\bea
\pi^+\pi^-\to \pi^+\pi^-&:&\quad
A_{--}^{++}=\frac{a_{--}^{++}+ik_0^0[(a_{-0}^{+0})^2-
a_{--}^{++}a_{00}^{00}]}{1-ik_-^+a_{--}^{++}-ik_0^0a_{00}^{00}+
k_0^0k_-^+[-a_{00}^{00}a_{--}^{++}+
(a_{-0}^{0+})^2]}\, ,\nn\\
\pi^0\pi^0\to \pi^+\pi^-&:&\quad
 A_{0-}^{0+}=\frac{a_{0-}^{0+}}{1-ik_-^+a_{--}^{++}-ik_0^0a_{00}^{00}+
k_0^0k_-^+[-a_{00}^{00}a_{--}^{++}+(a_{0-}^{0+})^2]}\, , \nn\\
\pi^0\pi^0\to \pi^0\pi^0&:&\quad
A_{00}^{00}=\frac{a_{00}^{00}+ik_-^+[(a_{-0}^{+0})^2-
a_{--}^{++}a_{00}^{00}]}{1-ik_-^+a_{--}^{++}-ik_0^0a_{00}^{00}+
k_0^0k_-^+[-a_{00}^{00}a_{--}^{++}+
(a_{-0}^{0+})^2]}\, ,\nn\\
{\rm with}&\quad& k_-^+ =\sqrt{\frac s4 -\mu^2_{\pi^+}}  \equiv k,\quad
 k_0^0= \frac12 \sqrt{\frac s4 -\mu^2_{\pi^0}}=
\frac12 \sqrt{k^2+\Delta^2}\, .
\label{AC1}
\eea
Here
$\Delta^2=\mu^2_{\pi^+}-\mu^2_{\pi^0}\simeq 0.07\mu^2_{\pi^+}$.
 The factor $1/2$ in $k_0^0$  arises due to the identity of pions in the
$\pi^0\pi^0$ state.

We impose on the scattering length values the standard isotopic
relations:
 \bea
&& a_{--}^{++}=\frac23a_0(s)+\frac13a_2(s), \nn\\
&& a_{-0}^{+0}=-\frac23a_0(s)+\frac23a_2(s), \nn\\
&& a_{00}^{00}=2a_{--}^{++}+a_{-0}^{+0}= \frac23a_0(s)+\frac43a_2(s)\,
.
\label{AC2}
\eea
Then at large $k^2$, when $k^2>>\Delta^2$, the unitary amplitudes of
 eq. (\ref{AC1}) obey the isotopic relations:
\bea
 A_{--}^{++}&=&\frac{\frac23a_0(s)}{1-ika_0(s)}+
\frac{\frac13a_2(s)}{1-ika_2(s)},\nn\\
 A_{-0}^{+0}&=&\frac{-\frac23a_0(s)}{1-ika_0(s)}+
\frac{\frac23a_2(s)}{1-ika_2(s)}, \nn\\
 A_{00}^{00}&=&\frac{\frac23a_0(s)}{1-ika_0(s)}+
\frac{\frac43a_2(s)}{1-ika_2(s)}.
\eea
 The ($I=0$)-amplitude and the corresponding $S$-matrix read:
\bea &&\frac{a_0(s)}{1-ika_0(s)}= 2A_{--}^{++}-\frac12A_{00}^{00}=
 A_{--}^{++}-\frac12A_{-0}^{+0}, \nn\\
&& \exp[2i\delta_0^0(s)]=\frac{A_{--}^{++}
-\frac12A_{-0}^{+0}}{(A_{--}^{++}-\frac12A_{-0}^{+0})^*}
=\frac{A_{--}^{++}
-\frac12A_{-0}^{+0}}{(2A_{--}^{++}-\frac12A_{00}^{00})^*}=
\frac{2A_{--}^{++}
-\frac12A_{00}^{00}}{(2A_{--}^{++}-\frac12A_{00}^{00})^*}.
\eea
 In the $K^+\rightarrow e^+\nu(\pi^+\pi^-)$ decay the  $S$-wave
pions are $I=0$ states, and the amplitude can be written as follows:
\bea
&& A\bigg(K^+\rightarrow e^+\nu(\pi^+\pi^-)_{I=0,
S-wave}\bigg)=
\lambda[1-ik_0^0A_{0-}^{0+}+ik_-^+A_{--}^{++}]= \nn \\
&&\lambda
[\frac{1-ik_0^0a_{00}^{00}-ik_0^0a_{0-}^{0+}}{1-ik_-^+a_{--}^{++}-
ik_0^0a_{00}^{00}+k_0^0k_-^+[-a_{00}^{00}a_{--}^{++}+(a_{0-}^{0+})^2]}]
\eea
 Here the first term, $\lambda$, is a direct production amplitude while
the second and third terms take into account pion rescatterings.

At large pion relative momentum, when $k^2>>\Delta^2$, we have:
\bea
&&A\bigg(K^+\rightarrow e^+\nu(\pi^+\pi^-)_{I=0, S-wave}
\bigg)_{k^2>>\Delta^2}=
\lambda \frac{1}{1-ika_0(s)}\, .
\eea
 Recall that the factor $(1-ika_0(s))^{-1}$ is due to rescatterings of pions
in the $I=0$ state.


\begin{thebibliography}{99}

\bibitem{PDG} W.-M. Yao {\it et al.}, PDG, J. Phys. G: Nucl. Part.
Phys. {\bf 33}, 1 (2006).
\bibitem{Ablikim} M. Ablikim et al. (BES Collab.) Phys. Lett. B{\bf}
645, 19 (2007).
\bibitem{Garcia} R. Garcia-Martin, J.R. Pelaez, F.J. Yndurain,
 Phys. Rev. D {\bf 76} 074034 (2007).
\bibitem{Bugg:2006gc}
  D.~V.~Bugg,
  J.\ Phys.\ G {\bf 34}, 151 (2007)
  [arXiv:hep-ph/0608081].
\bibitem{Caprini:2005zr}
  I.~Caprini, G.~Colangelo and H.~Leutwyler,
  Phys.\ Rev.\ Lett.\  {\bf 96}, 132001 (2006)
  [arXiv:hep-ph/0512364].
\bibitem{km} V.V. Anisovich and A.V. Sarantsev, Eur. Phys. J.
A~{\bf16}, 229 (2003);\\
 V.V. Anisovich, A.A. Kondashov, Yu.D. Prokoshkin, S.A.
Sadovsky, and A.V. Sarantsev, Yad.  Fiz.  {\bf 60}, 1489 (2000) [Phys.
Atom. Nucl. {\bf 60}, 1410 (2000)];\\
V.V. Anisovich and A.V. Sarantsev, Phys. Lett. B {\bf 382}, 429 (1996).

\bibitem{kmR} V.V. Anisovich and A.V. Sarantsev, Int. J. Mod. Phys.
A{\bf 24}, 2481, (2009);\\
V.V. Anisovich and A.V. Sarantsev,  Yad. Fiz. {\bf 72}, 1950 (2009) [Phys. Atom.
Nucl. {\bf 72}, 1889  (2009)];\\
V.V. Anisovich and A.V. Sarantsev,  Yad. Fiz. {\bf 72}, 1981 (2009) [Phys. Atom.
Nucl. {\bf 72}, 1920 (2009)].

\bibitem{APS}
V.V. Anisovich, Yu.D. Prokoshkin, and A.V. Sarantsev, Phys. Lett. B
{\bf 389}, 388 (1996).
\bibitem{book3} A.V. Anisovich, V.V. Anisovich, M.A. Matveev,
   V.A. Nikonov, J. Nyiri and A.V. Sarantsev, {\it Mesons and Baryons},
World Scientific, Singapore (2008).
\bibitem{AN} V.V. Anisovich and V.A. Nikonov, Eur. Phys. J. A{\bf 8},
 401 (2000).
\bibitem{GarciaMartin:2011cn}
  R.~Garcia-Martin, R.~Kaminski, J.~R.~Pelaez, J.~R.~de Elvira and F.~J.~Yndurain,
  arXiv:1102.2183 [hep-ph].

\bibitem{glueball} A.V. Anisovich, V.V. Anisovich,
Yu.D. Prokoshkin, and A.V. Sarantsev, Zeit. Phys. A {\bf 357}, 123
(1997).
\bibitem{ZPhys} A.V. Anisovich, V.V. Anisovich, and A.V. Sarantsev,
Phys. Lett. B {\bf 395}, 123 (1997);
Zeit. Phys. A {\bf 359}, 173 (1997).



\bibitem{Shapiro}I.S. Shapiro, Nucl. Phys. A {\bf 122}, 645 (1968).
\bibitem{Okun}
I.Yu. Kobzarev, N.N. Nikolaev, and L.B. Okun,
Sov. J. Nucl. Phys. {\bf 10}, 499 (1970).
\bibitem{Stodolsky}
L. Stodolsky, Phys. Rev. D {\bf 1}, 2683 (1970).
\bibitem{ABS}
V.V. Anisovich, D.V. Bugg, and A.V. Sarantsev, Phys. Rev. D {\bf
58}:111503 (1998).

\bibitem{ufn} V.V. Anisovich, UFN {\bf 168}, 481 (1998)
[Physics-Uspekhi {\bf 41}, 419 (1998)].

\bibitem{SI-qq} V.V. Anisovich, L.G. Dakhno,  M.A. Matveev, V.A. Nikonov,
and A.V. Sarantsev,  Yad. Fiz. {\bf 70}, 480 (2007) [Phys. Atom.
Nucl. {\bf 70}, 450 (2007)]; hep-ph/0511109.

\bibitem{SI-bb} V.V. Anisovich, L.G. Dakhno,  M.A. Matveev, V.A. Nikonov,
and A. V. Sarantsev,  Yad. Fiz. {\bf 70}, 68 (2007) [Phys. Atom.
Nucl. {\bf 70}, 63 (2007)]; hep-ph/0510410.

\bibitem{SI-cc} V.V. Anisovich,  L.G. Dakhno,  M.A. Matveev, V.A. Nikonov,
and A.V. Sarantsev,  Yad. Fiz. {\bf 70}, 392 (2007) [Phys. Atom.
Nucl. {\bf 70}, 364 (2007)]; hep-ph/0511105.

\bibitem{gpi} A.V. Anisovich,
V.V. Anisovich, L.G. Dakhno,  M.A. Matveev, V.A. Nikonov,
and A. V. Sarantsev,  J. Phys. G {\bf 37}:025004 (2010);\\
V.V. Anisovich, L.G. Dakhno,  M.A. Matveev, V.A. Nikonov,
and A. V. Sarantsev, Yad. Fiz. {\bf 73}, 488 (2010) [Phys. Atom.
Nucl. {\bf 73}, 462 (2010)]; hep-ph/0901.4854.


\bibitem{Salpeter} E. Salpeter and H.A. Bethe, Phys. Rev. {\bf 84},
1232 (1951).

\bibitem{syst}
A.V. Anisovich, V.V. Anisovich, and A.V.~Sarantsev, Phys. Rev.
D~{\bf 62}, 051502(R) (2000).





\bibitem{Gribov} J. Nyiri (ed.), '{\it The Gribov Theory of Quark
Confinement}', World Scientific, Singapore (2001).




\bibitem{W800P} S.D. Protopopescu {\it et al.},
Phys. Rev. D {\bf 7}, 1279  (1973).
\bibitem{W800E} P. Estabrooks, Phys. Rev. D {\bf 19}, 2678 (1979).
\bibitem{W800A} K.L. Au, D. Morgan and M.R. Pennington,
Phys. Rev. D {\bf 35}, 1633 (1987).
\bibitem{W800Ishida} S. Ishida {\it et al.}, Prog. Theor. Phys. {\bf 98},
 1005 (1997).

\bibitem{Aloisio} A. Aloisio, et al. Phys. Lett.,
 {\bf B538}, 21 (2002).

\bibitem{Gasser} G. Colangelo, J. Gasser, H. Leutwyler, Nucl. Phys.
 {\bf B603}, 125 (2001).

\bibitem{ANS-content} V.V. Anisovich, V.A. Nikonov and A.V.~Sarantsev,
Yad. Fiz. {\bf 66}, 772 (2003) [Phys. Atom.
Nucl. {\bf 66}, 741 (2003)].



\bibitem{WMandelstam} G.F. Chew and S. Mandelstam,
 Phys. Rev.  {\bf 119}, 467 (1960).

\bibitem{WChew} G.F. Chew, {\em The Analytic S-Matrix}, W.A. Benjamin, New
York, 1966.

\bibitem{Pislak} S. Pislak, et al. Phys. Rev. Lett.,
 {\bf 87}, 221801 (2001).

\bibitem{WBasdevant} J.L. Basdevant, C.D. Frogatt and J.L. Petersen,
Phys. Lett. B {\bf 41}, 178 (1972).
\bibitem{WZinn}
D. Iagolnitzer, J. Justin, and J.B. Zuber, Nucl. Phys. B {\bf 60},
 233 (1973).
\bibitem{WBugg} B.S. Zou and D.V. Bugg, Phys. Rev. D {\bf 48}, (1994)
R3942; {\it ibid}, D {\bf 50}, 591 (1994).
\bibitem{WSpeth} G. Janssen, B.C. Pearce, K. Holinde, and J. Speth,
Phys. Rev. D {\bf 52}, 2690 (1995).



%
%
\end{thebibliography}
\end{document}